\documentclass[12pt]{article}
\usepackage{amsmath}
\def\Box{{\hbox{$\sqcup$}\llap{\hbox{$\sqcap$}}}}
\renewcommand{\theequation}{\thesection.\arabic{equation}}

\begin{document}

\title{\bf \LARGE {A self-tuning mechanism in $\mathbf{(3+p)}$d gravity-scalar theory}}

\author{ \\Eun Kyung Park and Pyung Seong Kwon\footnote{E-mail : bskwon@ks.ac.kr}\\ Department of
Physics, Kyungsung University, Pusan 608-736, Korea}
\date{}

\maketitle

\thispagestyle{empty}

\vskip 0.5cm
\begin{abstract}
We present a new type of self-tuning mechanism for ($3+p$)d brane
world models in the framework of gravity-scalar theory. This new
type of self-tuning mechanism exhibits a remarkable feature. In
the limit $g_s \rightarrow 0$, $g_s$ being the string coupling,
the geometry of bulk spacetime remains virtually unchanged by an
introduction of the Standard Model(SM)-brane, and consequently it
is virtually unaffected by quantum fluctuations of SM fields with
support on the SM-brane. Such a feature can be obtained by
introducing Neveu-Schwarz(NS)-brane as a background brane on which
our SM-brane is to be set. Indeed, field equations naturally
suggest the existence of the background NS-brane. Among the given
such models, of the most interest is the case with $\Lambda=0$,
where $\Lambda$ represents the bulk cosmological constant. This
model contains a pair of coincident branes (of the SM- and the
NS-branes), one of which is a codimension-2 brane placed at the
origin of 2d transverse space ($\equiv \Sigma_2$), another a
codimension-1 brane placed at the edge of $\Sigma_2$. These two
branes are (anti) T-duals of each other, and one of them may be
identified as our SM-brane plus the background NS-brane. In the
presence of the background NS-brane (and in the absence of
$\Lambda$), the 2d transverse space $\Sigma_2$ becomes an orbifold
$R_2 /Z_n$ with an appropriate deficit angle. But this is only
possible if the ($3+p$)d Planck scale $M_{3+p}$ and the string
scale $M_s$($\equiv 1/\sqrt{\alpha^{\prime}}$) are of the same
order, which accords with the hierarchy assumption \cite{1,2,3}
that the electroweak scale $m_{EW}$ is the only short distance
scale existing in nature.
\end{abstract}

\vskip 1.0cm
\medskip
\begin{center}
\end{center}

\newpage
\setcounter{page}{1}

\baselineskip 6.0mm

\section{Introduction}

 Inspired by the fact that the SM fields in string theory should be
confined to the D-brane \cite{1,2}, it was proposed that our
universe may be a topological wall (3-brane) imbedded in a higher
dimensional spacetime (bulk) \cite{3}. This "brane world" scenario
has drawn considerable attention over the last few years since it
provides a mechanism for solving the longstanding puzzles such as
the cosmological constant problem \cite{4} or the hierarchy
problem \cite{2}. For instance, a 5d model was presented in
\cite{5} where the SM vacuum energy (or the brane tension) "warps"
only the 5d bulk spacetime and does not affect the geometry of the
brane itself, so the brane is kept flat regardless of the value of
the brane tension. In this model, the desired $TeV$ physical mass
scale can be obtained from the fundamental Planck scale $\sim
10^{19}GeV$ through an exponential hierarchy generated by an
exponential "warp" factor. Similar models also have been
considered in 6d \cite{6}. These models provide a nice way to
address the hierarchy problem, but they require fine-tuning
between brane and bulk parameters in order to admit static
solutions which are flat in the 4d brane world sector. Since these
models are not free from the fine-tuning, a separate discussion
would be necessary in order to meet the cosmological constant
problem.

In this context, models without fine-tuning (or self-tuning
models) have been proposed both in 5d \cite{7,8,9} and 6d
\cite{10,11,12}. In particular, in \cite{13} it was argued that
the self-tuning of the cosmological constant is generic in 5d
theories with no more than two branes coupled to a scalar field as
well as gravity. The idea of self-tuning mechanism in 5d is that
if in some cases the number of free parameters in the bulk
solution is greater than the number of matching conditions
(including orbifold boundary conditions), then one can find
solutions where the brane tension can take any arbitrary value
without changing geometry of the brane, and quantum corrections to
the brane tension do not disturb the flatness of the brane.
However, the presence of the bulk cosmological constant $\Lambda$
leads to a reintroduction of the fine-tuning between $\Lambda$ and
the brane tension except for a particular case \cite{8}. Also, 5d
models generally involve a naked singularity from a finite proper
distance from the brane \cite{9,13}.

As mentioned above, self-tuning models also have been proposed in
6d. The self-tuning mechanism in 6d is different from that in 5d.
Generally, 6d solutions have a desirable property that the brane
tension does not affect the 4d effective cosmological constant; it
only induces a deficit angle in the transverse space. The above 6d
models are worth noticing in this respect. But still, certain
kinds of fine-tunings are necessary in these models. For instance,
a fine-tuning of bulk parameters is needed in \cite{11} to obtain
a sufficiently small value for the 4d cosmological constant, or a
flux quantization causes a reintroduction of the fine-tuning
between brane and bulk parameters in \cite{12}. Besides this,
there have been some claims \cite{14} that these 6d models are not
really the self-tuning models; they are indeed fine-tuning models.
(Further conventional works in 6d can be found in \cite{15}.)

In this paper, along this line of studies, we will consider
($3+p$)d brane world solutions (and corresponding brane world
models) in the framework of gravity-scalar theory. The solutions
show that the ($3+p$)d gravity-scalar action admits a new type of
self-tuning mechanism distinguished from the conventional ones.
Namely, the geometry of bulk spacetime remains virtually
unaffected by the quantum fluctuations of SM fields with support
on the SM-brane in the limit as the string coupling $g_s$ goes to
zero. Such a feature involves an introduction of a background
NS-brane, whose existence is naturally suggested by field
equations.

Historically, the theory with background NS-branes in the limit
$g_s \rightarrow 0$ is not new. It is believed that in the limit
$g_s \rightarrow 0$ NS5-branes of the type II string theory
decouple from bulk modes (including gravity), and this decoupled
theory of NS5-branes is known as "Little String
Theory"(LST)\cite{16}. For this reason the discussion of this
paper may be partially regarded as an analogue of LST, and
consequently the limit $g_s \rightarrow 0$ as an analogue of the
decoupling limit of LST. Such an implication in LST is briefly
discussed in Sec.6 and Sec.7.

Besides this, the self-tuning mechanism of this paper is naturally
connected with the hierarchy problem. In the presence of the
background NS-brane (and in the absence of $\Lambda$), the 2d
transverse space becomes an orbifold $R_2 / Z_n$ with a deficit
angle $\delta = 2\pi (1-\beta)$ with $\beta \sim (M_s
/M_{3+p})^{1+p}$, where $M_s$ is the string scale, while $M_{3+p}$
the $(3+p)$d Planck scale. So in order that $\beta$ becomes of
order one, $M_s$ and $M_{3+p}$ must be of the same order, which
naturally accords with the hierarchy conjecture that there may
exist only one fundamental short distance scale (which is
presumably the electroweak scale) in nature. We will see this in
Sec.10.

\section{Reduced action}

We start with a ($3+p$)d action given by\footnote{Instead of the
last term of (2.1), it is also possible to consider more
generalized terms like $(-1/2) \mathop{\Sigma}\limits^{2}_{i=1}
e^{\sigma_i} (\nabla \Psi_i)^2$, where $\sigma_i$ is any dilatonic
scalar other than $\Phi$. The whole discussion of this paper still
holds for this generalized Lagrangian as can be inferred from the
argument of Sec.5.}
\begin{equation}
I= \frac{1}{2\kappa^2} \int d^{3+p}X \sqrt{-G} \Big[e^{-2\Phi}
\big[ R + 4 (\nabla \Phi)^2 \big] - e^{\alpha \Phi} \Lambda -
\frac{1}{2} \mathop{\Sigma}\limits^{2}_{i=1} (\nabla \Psi_{i})^2
\Big]\,\,,
\end{equation}
where $2\kappa ^2 = 1/2{M_{3+p}^{1+p}}$ in terms of the ($3+p$)d
Planck scale $M_{3+p}$, and $\Phi$ is the ($3+p$)d dilaton. The
cosmological constant term includes a factor $e^{\alpha \Phi}$,
and where the case $\alpha =-2$ is of particular interest because
it corresponds to the string theory. The action (2.1) also
contains two scalar fields $\Psi_1$ and $\Psi_2$, which will play
an essential role in our self-tuning mechanism. To obtain
equations of motion from (2.1) it is convenient to follow the
lines of \cite{17}. We take the ($3+p$)d metric to be of the form
\begin{equation}
ds{_{3+p}^2} = e^{A(r)} d\hat{s}{_3^2} + e^{B(r)} d\vec{x}{_p^2}
\end{equation}
where $d\vec{x}{_p^2}=dx{_1^2} + \cdots + dx{_p^2}$, the line
element of the $p$-brane, while
\begin{equation}
d\hat{s}{_3^2}  = -N^2 (r)dt^2 + \frac{dr^2}{f^2(r)} + R^2 (r)
d\theta^2 \equiv \hat{g}_{ab} dy^a dy^b
\end{equation}
represents a 3d subspace($\equiv \Sigma_3$) with coordinates $y^a
\equiv (t, r, \theta)$. In (2.3), $\theta$ is an angular variable
with $0 \leq \theta \leq 2 \pi$, and $(r, \theta)(\equiv \vec{r})$
are polar coordinates representing 2d transverse space($\equiv
\Sigma_2$). At this point we implicitly assume that the geometry
of $\Sigma_2$ is locally $R_2$ around $\vec{r} =0$.

The metric (2.2) is of the most general form with an
$\large{O}(p)$ symmetry on the brane sector. In fact the field
degrees of freedom of the metric (2.2) (together with (2.3)) are
redundant. For instance the factor $e^A$ could be absorbed into
$d\hat{s}{_3^2}$, but it has been introduced for the later use.
$f(r)$ is also a redundant degree of freedom associated with a
coordinate transformation $r \rightarrow \tilde{r}=F(r)$. Now
notice that the scale factors in (2.2) and (2.3) all depend only
on $r$. We will also assume that the scalar fields $\Phi$,
$\Psi_1$, and $\Psi_2$ are all functions of $r$ alone (namely, we
are considering rotationally symmetric $p$-brane solutions). Since
the fields do not depend on the coordinates $x_i$ along the
$p$-brane, the ($3+p$)d action (2.1) can be reduced to a 3d action
defined on the subspace $\Sigma_3$:
\begin{eqnarray}
I_{red} = \frac{1}{2\kappa^2} \int  d^3 y \sqrt{-
\det|\hat{g}_{ab}|}\,\,e^{-2\phi} \Big[\,\,\hat{R}\,-\,
e^{A}e^{(\alpha+2)\Phi}\Lambda \,-\, \frac{1}{4}(\partial A)^2
\,-\,\frac{p}{4}(\partial B)^2 \,+ \, 4 (\partial \phi)^2 \nonumber \\
-\, 2(\partial \phi)(\partial A) \, -\,  \frac{1}{2} e^{2 \Phi}
\mathop{\Sigma}\limits^{2}_{i=1} (\partial \Psi_{i})^2
\,\Big]\,\,,~~~~~~~~~~~~~~~~~~~~
\end{eqnarray}
where $\phi$ is the 3d effective dilaton defined by
\begin{equation}
\phi = \Phi - \frac{1}{4}A - \frac{p}{4}B \,\,,
\end{equation}
and $\hat{R}$ is the 3d Ricci-scalar obtained from $\hat{g}_{ab}$.
The indices in (2.4) are all raised or lowered with $\hat{g}_{ab}$.
If we choose (recall that $e^A$ was a surplus degree of freedom)
\begin{equation}
A=4\Phi - pB \,\,,
\end{equation}
(2.4) reduces to
\begin{eqnarray}
I_{red} = \frac{1}{2\kappa^2} \int  d^3 y \sqrt{-
\det|\hat{g}_{ab}|}\,\,\Big[\,\,\hat{R}\,-\,e^{(\alpha+6)\Phi -pB
}\Lambda \,-\,4(\partial \Phi)^2 + 2p(\partial \Phi) (\partial B) \nonumber \\
-\, \frac{p(p+1)}{4}(\partial B)^2  \,-\, \frac{1}{2}
e^{2\Phi}\mathop{\Sigma}\limits^{2}_{i=1} (\partial \Psi_{i})^2
\,\Big]\,\,,~~~~~~~~~~~~~~~~~~~~
\end{eqnarray}
where the 3d dilaton $\phi$ is absent and the kinetic term for
$\hat{g}_{ab}$ has the standard Einstein form without coupled to
$\phi$. So the first term of (2.7) is a standard Hilbert-Einstein
action for the 3d gravity $\hat{g}_{ab}$, while the remaining
terms constitute an effective matter action that gives a
contribution to the energy-stress tensor $T_{ab}$ of the 3d Einstein
equations for $\hat{g}_{ab}$.

\setcounter{equation}{0}
\section{Codimension-2 brane and Field equations with
{$\mathbf{\Psi_i =0}$}}

The bulk equations of motion off the brane can be obtained from
the reduced action in (2.7). But in order to include the effect of
the brane we need to introduce a brane action. In this section we
start the discussion with an assumption that we have a
codimension-2 brane at $\vec{r} =0$. Also, we assume that the
scalar fields $\Psi_i$ are "turned off" ($\Psi_i =0$) at this
point. The action for a dilaton-coupled codimension-2 brane is
most generally given by
\begin{equation}
I_{cod-2} = - \int d^{p+1}x \sqrt{-det|g_{\mu\nu}|} \,\,V_p(\Phi)
\,\,,
\end{equation}
where $V(\Phi)$ is an arbitrary functional of $\Phi$, and
$g_{\mu\nu}$ is a pullback of $G_{AB}$ to the ($p+1$)d brane
world:
\begin{eqnarray}
ds{_4^2} = G_{\mu\nu} dX^\mu dX^\nu |_{\vec{r}=0}
~~~~~~~~~~~~~~~~~~~~~~~~~~~~~~~~~~~~\nonumber \\ = - e^A N^2 dt^2
+e^B \big(dx{_1^2} + \cdots +dx{_p^2} \big)\Big|_{\vec{r}=0}
~~~~~~~~~ \nonumber\\ \equiv g_{\mu\nu} dx^\mu dx^\nu
\,\,.~~~~~~~~~~~~~~~~~~~~~~~~~~~~~~~~~~~~~~~~~~
\end{eqnarray}
Upon using (2.6) and (3.2), the action $I_{cod-2}$  can be converted
into
\begin{equation}
I_{cod-2} = - \int \limits^{}_{\Sigma_3} d^3y \sqrt{-det |
\hat{g}_{ab}|}\,\,e^{2\Phi} \, V_p(\Phi)\,\, \delta^2(\vec{r})
\,\,,
\end{equation}
where the 2d delta-function $\delta^2(\vec{r})$ has been normalized by
\begin{equation}
 \int \limits^{}_{\Sigma_2} d^2 \vec{r} \sqrt{\hat{g}_2}\,\, \delta^2(\vec{r}) =1 \,\,,
\end{equation}
where $\hat{g}_2$ represents the determinant of the 2d metric
defined on $\Sigma_2$. As mentioned above we will begin our
discussion with a codimension-2 brane placed at $\vec{r}=0$. But
later we will introduce a codimension-1 brane (in the case
$\Lambda=0$) at the edge of $\Sigma_2$, which becomes a T-dual of
the codimension-2 brane at $\vec{r}=0$. Such a codimension-1 brane
appears as a circle of certain radius, while the codimension-2
brane simply as a point on $\Sigma_2$.

The field equations with a codimension-2 brane are now obtained
from the total action $I_{total} \equiv I_{red}+I_{cod-2}$. In
covariant form they are written as

\vskip 0.1cm \hspace{4cm}$\rm {(a)~}$ $\rm {3d~ Einstein~
equations:}$
\begin{equation}
\hat{R}_{ab} - \frac{1}{2} \hat{g}_{ab} \hat{R} = \kappa^2
(T_{ab}+t_{ab})\,\,,~~~~~(a,b=t, r, \theta)~~~~~~~~~~~~
\end{equation}
with
\begin{eqnarray}
T_{ab} = -\frac{1}{2\kappa^2}\hat{g}_{ab}\, e^{(\alpha +6)\Phi -pB} \Lambda +
\frac{4}{\kappa^2}\Big[(\partial_a \Phi)(\partial_b
\Phi)-\frac{1}{2}\hat{g}_{ab}(\partial \Phi)^2
\Big]-\frac{2p}{\kappa^2}\Big[(\partial_a \Phi)(\partial_b
B)\nonumber\\-\frac{1}{2}\hat{g}_{ab}(\partial \Phi)(\partial
B)\Big] + \frac{p(p+1)}{4\kappa^2}\,\Big[(\partial_a B)(\partial_b
B)-\frac{1}{2} \hat{g}_{ab} (\partial B)^2 \Big] \,\,,~~~~~~~~~~
\end{eqnarray}
\begin{equation}
t_{ab}=- \hat{g}_{at} \hat{g}^{tt} \hat{g}_{tb}\, e^{2\Phi}
V_{p}(\Phi)\, \delta^2 (\vec{r}) \,\,,
\end{equation}
where $t_{ab}$ follows from the action $I_{cod-2}$ in (3.3).

\vskip 0.1cm
\hspace{3.5cm}$\rm {(b)~}$ $\rm {Equations~ for~ \Phi
~ and~ B:}$
\begin{equation}
\Box \Phi -\frac{p}{4} \Box B- \frac{(\alpha +6)}{8}e^{(\alpha +6)\Phi -pB} \Lambda =\frac{\kappa^2}{2}\,e^{2\Phi}
\Big[V_{p}(\Phi) + \frac{1}{2}\frac{\partial V_{p}(\Phi)}{\partial \Phi}\Big]
\,\delta^2 (\vec{r}) \,\,,
\end{equation}
\begin{equation}
\frac{(p+1)}{4} \Box B - \Box \Phi + \frac{1}{2} e^{(\alpha +6)\Phi -pB} \Lambda = 0 \,\,.
\end{equation}
Substituting (2.3) into the Einstein equations in (3.5) gives the
following three independent equations of motion:
\begin{equation}
 N(fR^{\prime})^{\prime} + \frac{1}{2} \frac{NR}{f} e^{(\alpha
+6)\Phi -pB} \Lambda +
 NfR \mathcal{H} \,=\,-\kappa^2 \frac{NR}{f} e^{2\Phi} \, V_p(\Phi)\, \delta^2(\vec{r}) \,\,,
\end{equation}
\begin{equation}
 N^{\prime}fR^{\prime} + \frac{1}{2} \frac{NR}{f} e^{(\alpha +6)\Phi -pB} \Lambda -
 NfR \mathcal{H} \,=\, 0,
\end{equation}
\begin{equation}
(N^{\prime}f)^{\prime}R + \frac{1}{2} \frac{NR}{f} e^{(\alpha +6)\Phi
-pB} \Lambda + NfR \mathcal{H} \,=\, 0 \,\,,
\end{equation}
where $\mathcal{H} \equiv 2 {\Phi^{\prime}}^2 -p\Phi^{\prime} B^{\prime} + \frac{p(p+1)}{8} {B^{\prime}}^2$,
and the "prime" denotes the derivative with respect to $r$.
Similarly, from (3.8) and (3.9) one obtains
\begin{eqnarray}
(NfR\Phi^{\prime})^{\prime} - \frac{[(\alpha+ 2)(p+1)+4
]}{8} \frac{NR}{f} e^{(\alpha +6 )\Phi -pB}
\Lambda \nonumber\\
=\, \frac{(p+1)}{2}\kappa^2 \frac{NR}{f} e^{2\Phi} \,\Big[
V_{p}(\Phi) + \frac{1}{2}\frac{\partial V_{p}(\Phi)}{\partial
\Phi} \Big] \,\delta^2 (\vec{r}) \,\,,
\end{eqnarray}
\begin{equation}
(NfRB^{\prime})^{\prime} -\frac{(\alpha +2)}{2} \frac{NR}{f}
e^{(\alpha +6 )\Phi -pB} \Lambda\,=\, 2\kappa^2 \frac{NR}{f}
e^{2\Phi} \Big[ V_{p}(\Phi) + \frac{1}{2}\frac{\partial
V_{p}(\Phi)}{\partial \Phi}\Big] \,\delta^2 (\vec{r})\,\,.
\end{equation}

Equations (3.10)-(3.14) constitute a complete set of linearly
independent equations of motion. Now we set
\begin{equation}
N \equiv e^{-2\Phi +\frac{(p+1)}{2}B}\, h \,\,,
\end{equation}
which (together with (2.6)) converts (2.2) into
\begin{equation}
ds{_{3+p}^2}  = e^{4\Phi -pB} \Big(\frac{dr^2}{f^2} + R^2
d\theta^2 \Big) + e^B \Big(-h^2 dt^2 + d \vec{x}{_p^2} \,\Big)
\,\,.
\end{equation}
For $h=f$ (and with $\Phi = (p-1)B/4$), (3.16) takes the usual
form of the black brane. But in this paper we are interested in
the case $h=1$, i.e., the solution preserving ($p+1$)d
Poincar\'{e} invariance. By substituting (3.15) into (3.11)-(3.14)
and comparing with one another, one obtains
\begin{equation}
 (NfR\frac{h^\prime}{h})^\prime \,=\,0\,\,,
\end{equation}
which shows that the ($p+1$)d Poincar\'{e} invariance $(h=1)$ is
automatically preserved. In what follows we will set
\begin{equation}
h=1\,\,,
\end{equation}
then
\begin{equation}
N \longrightarrow \xi \equiv e^{-2\Phi + \frac{(p+1)}{2}B} \,\,.
\end{equation}

Due to (3.17), Eqs. (3.10)-(3.14) are no longer linearly
independent; only four of them are. Omitting (3.12), and choosing
$f(r)$ as
\begin{equation}
f \,=\, \frac{r}{\xi R} \,\,,
\end{equation}
one finds that the field equations reduce to the
following set of four linearly independent equations:
\begin{equation}
\nabla^2 \ln R + \Lambda\,\psi = -\kappa^2 C_1 \, \delta^2
(\vec{r})\,\,,
\end{equation}

\begin{equation}
\nabla^2 \Phi -\frac{[(\alpha +2)(p+1)+4]}{8} \Lambda\,\psi =
\frac{(p+1)}{2}\kappa^2 C_2 \,\delta^2 (\vec{r})\,\,,
\end{equation}
\begin{equation}
\nabla^2 B -\frac{(\alpha +2)}{2} \Lambda\,\psi = 2\kappa^2 C_2 \,
\delta^2 (\vec{r})\,\,,
\end{equation}
\begin{eqnarray}
- 2\big(r\frac{d\ln \xi}{dr}\big)\big(r\frac{d\ln R}{dr}\big)+2\big[
2\big(r\frac{d\Phi}{dr}\big)^2 - p
\big(r\frac{d\Phi}{dr}\big)\big(r\frac{dB}{dr}\big)+\frac{p(p+1)}{8}\big(r\frac{dB}{dr}\big)^2
\big] \nonumber\\ = \Lambda \psi r^2 \,\,,
~~~~~~~~~~~~~~~~~~~~~~~~~~~~
\end{eqnarray}
where $\nabla^2$ is the flat space Laplacian $\nabla^2 \equiv
(1/r)(d/dr)(rd/dr)$ (so $\delta^2 (\vec{r})$ is now normalized by
$\int r dr d\theta \delta^2 (\vec{r})=1$), while $\psi$ and $C_i$
are defined, respectively, by
\begin{equation}
\psi = \frac{\xi^2 R^2 e^{(\alpha +6)\Phi -pB}}{r^2}
\,=\,\frac{R^2 e^{(\alpha +2)\Phi +B}}{r^2} \,\,,
\end{equation}
and
\begin{equation}
C_1 = e^{(p+1)B/2} V_{p}(\Phi)\Big|_{\vec{r}=0}\,\,,~~~~~ C_2 =
e^{(p+1)B/2} \Big[ V_{p}(\Phi) + \frac{1}{2}\frac{\partial
V_{p}(\Phi)}{\partial \Phi} \Big]\Big|_{\vec{r}=0} \,\,.
\end{equation}
By inspecting (3.25) together with (3.21)-(3.23) one finds that $\psi$
must satisfy
\begin{equation}
\nabla^2 \ln \psi + m\Lambda\,\psi = \kappa^2 \Big[ 2(C_2 -C_1 )+
\frac{(\alpha +2)(p+1)}{2}C_2 \Big] \delta^2 (\vec{r}) - 4\pi
\delta^2 (\vec{r})\,\,,
\end{equation}
where
\begin{equation}
m \equiv 2 - \frac{(\alpha +2)[(\alpha +2)(p+1)+8]}{8} \,\,,
\end{equation}
and similarly from (3.19), (3.22) and (3.23), we get
\begin{equation}
\nabla^2 \ln \xi + \Lambda\,\psi = 0 \,\,.
\end{equation}
Also in terms of $\psi$ and $\xi$ the metric (3.16) is rewritten as
\begin{equation}
ds^2_{3+p} = e^{-(\alpha +2)\Phi}\,\psi \,\Big(dr^2 +
\frac{r^2}{\xi^2}d\theta^2 \Big) + e^B \Big(-dt^2 + d
\vec{x}{_p^2} \Big) \,\,.
\end{equation}

\setcounter{equation}{0}
\section{Solution to field equations}

The solution to the differential equations in the previous section
can be readily found. Using (3.27), one can show that the most
general solution to the set of field equations (3.21)-(3.23)
and (3.29) takes the form
\begin{equation}
R=i_R \, \psi^{k_R} R_0 \,\,, ~~~~e^{\Phi} =i_\Phi \,
\psi^{k_\Phi}\,\,, ~~~e^B = i_B \, \psi^{k_B} \,\,, ~~~ \xi =
i_\xi \, \psi^{k_\xi} \,\,,
\end{equation}
where $R_0$ is a constant with length dimension one, and $k_M$ ($M
\equiv R, \Phi, B, \xi$) are given by
\begin{equation}
k_R = k_\xi = \frac{1}{m}\,\,,~~~k_\Phi = -\frac{[(\alpha
+2)(p+1)+4]}{8m} \,\,, ~~~ k_B = - \frac{(\alpha +2)}{2m} \,\,.
\end{equation}
Also $i_M$ are defined by
\begin{equation}
\nabla^2 \ln i_M = 2\pi (\alpha_M + 2k_M ) \, \delta^2 (\vec{r})
\equiv 2 \pi a_M \delta^2 (\vec{r})\,\,,
\end{equation}
where the constants $\alpha_M$ are given by
\begin{equation}
\alpha_R = - \frac{\kappa^2}{2\pi} \Big[ (1-\frac{2}{m})C_1
+\frac{2}{m}C_2 + \frac{(\alpha +2)(p+1)}{2m}C_2 \Big] \,\,,
\end{equation}
\begin{equation}
\alpha_\Phi = - \frac{\kappa^2}{2\pi} \Big[\frac{[(p+1)+1]}{m}C_2
 - \frac{[(\alpha +2)(p+1)+4]}{4m} C_1 \Big]\,\,,
\end{equation}
\begin{equation}
\alpha_B = - \frac{\kappa^2}{2\pi} \Big[ \frac{4}{m}C_2  -
\frac{(\alpha +2)}{m}(C_1 + C_2 ) \Big]\,\,,
\end{equation}
\begin{equation}
\alpha_\xi = - \frac{\kappa^2}{2\pi} \Big[\frac{2}{m}(C_1  - C_2 )
- \frac{(\alpha +2)(p+1)}{2m}C_2 \Big]\,\,.
\end{equation}
By (4.7), (3.27) can be rewritten as
\begin{equation}
\nabla^2 \ln \psi + m\Lambda\,\psi = - 2\pi \,(m \alpha_\xi + 2 )
\, \delta^2 (\vec{r}) \,\,,
\end{equation}
and in the case $\Lambda \neq 0$, the solution to (4.8) takes the
form \cite{18}
\begin{equation}
\psi (r)= \frac{\pm ({8 \gamma^2}/{m \Lambda})}{r^2 \big[
({r}/{r_0})^{\gamma} \pm ({r}/{r_0})^{-\gamma} \big]^2}\,\,,
~~~~~\Big(\pm \equiv \frac{\Lambda}{|\Lambda|}\Big) \,\,,
\end{equation}
where $r_0$ is an arbitrary integration constant, but $\gamma$ is
given in terms of $\alpha_\xi$ :
\begin{equation}
\gamma = - \frac{m}{2} \alpha_\xi \,\,.
\end{equation}
For $\Lambda = 0$, on the other hand, the solution to (4.8) can be
written in the form
\begin{equation}
\psi (r) = \frac{b_0}{r^2 ({r}/{r_0})^{-2 \gamma}}  \,\,,
\end{equation}
where $b_0$ is an arbitrary constant with length dimension two. In
(4.9) (i.e., in the case $\Lambda \neq 0$), the constant $\gamma$
must be positive. If $\gamma <0$, $\psi$ becomes $\psi \sim
r^{-(2\gamma +2)}$ as $r \rightarrow 0$, and therefore $\nabla^2
\ln \psi \sim -2 \pi \,(-m \alpha_\xi +2)\,\delta^2 (\vec{r})$,
which does not agree with (4.8), i.e., $\psi$ does not satisfy the
boundary condition at $\vec{r}=0$. In the case $\Lambda=0$,
however, $\gamma$ (and consequently $\alpha_\xi$) can be both
positive and negative. Finally, the solution to (4.3) (for both
$\Lambda \neq 0$ and $\Lambda =0$) is simply
\begin{equation}
i_M (r) = \Big(\frac{r}{\tilde r_0} \Big)^{a_M} \,\,, ~~~~~(a_M
\equiv \alpha_M + 2 k_M )\,\,,
\end{equation}
where $\tilde{r}_0$ is an arbitrary constant.

Though the solution (4.1) satisfies (3.21)-(3.23) and (3.29), we
still need for consistency to check whether it satisfies (3.24)
either. Substituting (4.1) (together with (4.9) (or (4.11)) and (4.12)) into
(3.24) gives two consistency conditions
\begin{equation}
\frac{2}{m}(\alpha_\xi +\alpha_R ) + \frac{[(\alpha +2)+4]}{m}
\alpha_\Phi - \frac{p}{m} \alpha_B =0 \,\,,
\end{equation}
and
\begin{equation}
-2 \alpha_\xi \alpha_R + 4 \alpha_\Phi^2 - 2p\alpha_\Phi \alpha_B
+ \frac{p(p+1)}{4} \alpha_B^2 = \frac{2}{m} \gamma^2 \,\,.
\end{equation}
Since
\begin{equation}
\alpha_\xi = -2 \alpha_\Phi + \frac{(p+1)}{2} \alpha_B
\end{equation}
from the definition of $\xi$ (and from (4.1) and (4.12)), the first
consistency condition (4.13) can be rewritten as
\begin{equation}
2 \alpha_R + \alpha_B + (\alpha +2) \alpha_\Phi =0 \,\,.
\end{equation}
(4.16), however, is not a constraint. Using (4.4)-(4.6), one can show
that (4.16) is identically satisfied from the definitions of
$\alpha_M$. The condition (4.14), however, imposes a restriction on
$\alpha_M$. Using (4.10), (4.15) and (4.16) one finds that (4.14) leads to a
condition
\begin{equation}
 l + \frac{(\alpha +2)}{2} \equiv \hat{l} = 0 \,\,,
\end{equation}
where $l$ is defined by
\begin{equation}
l \equiv \frac{\alpha_B}{\alpha_\xi} \,\,.
\end{equation}
In the case $\alpha =-2$ (or $m=2$), (4.17) reduces to
\begin{equation}
l=0 ~~~~~ \longleftrightarrow ~~~~~\alpha_B =0 \,\,,
\end{equation}
which in turn implies $C_2 =0$ (readers can check that $\hat{l}=0$
in the case $\alpha \neq -2$ also leads to $C_2 =0$), or
equivalently
\begin{equation}
\big[ V_{p}(\Phi) + \frac{1}{2} \frac{\partial
V_{p}(\Phi)}{\partial \Phi} \big]\Big|_{\vec{r}=0} =0 \,\,.
\end{equation}

In the weakly coupled case of the string theory $V_{p}(\Phi)$ is
expected to be a power series of the form
\begin{equation}
V_{p}(\Phi)\,=\, e^{\beta\Phi}
\mathop{\Sigma}\limits^{\infty}_{n=0} V_n e^{n\Phi}  \,\,, ~~~~~
(V_n = {\rm const.})
\end{equation}
(where $\beta =-1$ for the D$p$-brane) when the quantum
corrections to the brane tension are controlled by the dilaton. So
if the brane at $\vec{r}=0$ is a D$p$-brane\footnote{To be
precise, the objects discussed in the present paper are not the
real BPS-objects of the string theories, as the action does not
include the terms for charges (i.e., $n$-form fields). The terms
D$p$-brane and NS-brane are used in analogy to string theory.
However, we will proceed the discussion assuming as if they are
BPS-branes, whose charges are neglected as is often the case with
the usual brane world scenarios.} with $V_p (\Phi)$ given by
(4.21), $V_n$ must be highly fine-tuned (for a given value of the
string coupling $e^{\Phi}\big|_{\vec{r}=0}$) in order to satisfy
(4.20), and this can hardly be accepted. Above all, D-brane does
not satisfy (4.20) at the tree level.\footnote{The tree level form
$V_p (\Phi) =V_0 e^{-\Phi}$ of the D$p$-brane can satisfy (4.20)
if we take $\Phi(\vec{r}=0)=\infty$. But this is not consistent
with the weakly coupled case in which $e^{\Phi}$ should not blow
up on the D-brane.} For the NS-brane, however, $V_p (\Phi)$ takes
the form
\begin{equation}
V_{p}(\Phi)\,=\,V_0 e^{-2\Phi}\,\,, ~~~~~(V_0 = {\rm const.})\,\,,
\end{equation}
and (4.20) is immediately solved by (4.22) for arbitrary $V_0$.
This suggests that the codimension-2 brane at $\vec{r}=0$ with
$\Psi_i =0$ would be an NS-NS type $p-$brane whose tension is
given by (4.22). The type II and the heterotic string theories
admit such a brane called NS5-brane. NS$p-$brane with $p<5$ may
then be regarded as a dimensionally reduced NS5-brane; for
instance, NS3-brane can be taken to be a subsector of the
NS5-brane wrapped on 2d compact space other than $\Sigma_2$ (see
for instance \cite{19}). This NS$p-$brane serves as a "background
brane" on which the SM-brane (a D$p-$brane with SM-fields living
on it) is to be set.

\setcounter{equation}{0}
\section{Form invariant action}

So far we have considered the case where $\Psi_i$ are absent. Then
we ended up with a result that the codimension-2 brane at
$\vec{r}=0$ should be an NS-brane. The action (2.1) with $\Psi_i
=0$ (and with $\Lambda =0$ if necessary) precisely coincides with
the low energy (closed) string action with zero NS-NS 2-form
field. But the result is that this action does not seem to admit a
natural D-brane solution, which immediately gets us into trouble
because SM fields live on a D-brane. In this section we will show
that such a difficulty can be avoided if we allow nonzero $\Psi_i$
in the action.

Turning back to the reduced action (2.4), one can show that the
effect of introducing nonzero $\Psi_i$ is to make a certain shift
in $\Phi$, $A$ and $B$. Namely, in the case $\Lambda =0$, the
action with the field contents $\Phi$, $A$, $B$ and nonzero
$\Psi_i$ is equivalent to the action with the new fields
$\tilde{\Phi}$, $\tilde{A}$, $\tilde{B}$ and vanishing $\Psi_i$ :
\begin{equation}
I(\Phi, A, B)\Big|_{\Psi_i \neq 0} = I(\tilde{\Phi}, \tilde{A},
\tilde{B})\Big|_{\Psi_i = 0} \,\,,
\end{equation}
where $\tilde{\Phi}$, $\tilde{A}$, $\tilde{B}$ are given by
\begin{equation}
\tilde{\Phi}=\Phi + F_{\Phi}\,\,, ~~~~~\tilde{A}=A+4F_{\Phi}-pF_B
\,\,, ~~~~~ \tilde{B}= B + F_B \,\,,
\end{equation}
and $F_\Phi$ and $F_B$ are related with $\Psi_i$ by
\begin{equation}
\Big[ (4 \partial F_\Phi - p \partial F_B )+ (4 \partial \Phi - p
\partial B) \Big]^2 - (4 \partial \Phi - p
\partial B )^2 = 2 e^{2\Phi} (\partial \Psi_{1})^2 \,\,,
\nonumber
\end{equation}
\begin{equation}
(\partial F_B + \partial B)^2 - (\partial B)^2 = \frac{p}{2}\,
e^{2\Phi} (\partial \Psi_{2})^2 \,\,.
\end{equation}
In (5.3), $\Psi_1 = \Psi_2 =0$ implies $F_\Phi =F_B =0$ (we
exclude the trivial case $F_\Phi$, $F_B$ $=$ constant), which
means that $F_\Phi$ and $F_B$ are field redefinitions of $\Psi_1$
and $\Psi_2$. In addition to this, $F_\Phi$ and $F_B$ also depend
on $e^{\Phi}$; i.e., they are functionals of $\Psi_1$, $\Psi_2$
and $e^{\Phi}$. (5.1) also holds for the case $\Lambda \neq 0$,
but this time we have only one $\Psi_i (\equiv \Psi)$. That is,
$\tilde{\Phi}$, $\tilde{A}$ and $\tilde{B}$ are given by
\begin{equation}
\tilde{\Phi}=\Phi + \frac{p}{4}F\,\,, ~~~~~\tilde{A}=A-
\frac{(\alpha +2)p}{4}F\,\,, ~~~~~ \tilde{B}= B + \frac{(\alpha
+6)}{4}F \,\,,
\end{equation}
and $F$ is related with $\Psi$ by
\begin{eqnarray}
\Big[ \partial F + \frac{2}{(4-m)} \big(\partial B + \frac{(\alpha
+2)}{2} \partial \ln \xi \big) \Big]^2 - \Big[ \frac{2}{(4-m)}
\big(\partial B + \frac{(\alpha +2)}{2} \partial \ln \xi \big)
\Big]^2 \nonumber \\
= \frac{4}{(4-m)p} \, e^{2\Phi} (\partial \Psi)^2
\,\,.~~~~~~~~~~~~~~~~~~~~
\end{eqnarray}
We see that $\Psi =0$ implies $F=0$ as before, and especially for
the string theory ($\alpha = -2$), (5.5) reduces to
\begin{equation}
(\partial F + \partial B)^2 - (\partial B)^2 = \frac{2}{p} \,
e^{2\Phi} (\partial \Psi)^2 \,\,.
\end{equation}

Let us consider the field equations in the presence of nonzero
$\Psi_i$. (5.2) (or (5.4)) shows that by appropriate field
redefinitions $\Psi_i$ can be absorbed into $\tilde{\Phi}$,
$\tilde{A}$, $\tilde{B}$ (in the forms of $F_M$ ($M \equiv \Phi,
B$) or $F$), and do not manifest themselves in the action. So the
field equations (and consequently the solution) following from the
action with $\Psi_i \neq 0$ are expected to take precisely the
same form as those following from the action with $\Psi_i =0$
except that ($\Phi$, $A$, $B$) are replaced by ($\tilde{\Phi}$,
$\tilde{A}$, $\tilde{B}$). This implies that for $\Psi_i \neq 0$,
(4.20) should be replaced by
\begin{eqnarray}
\Big[ V_{p}(\tilde{\Phi})+\frac{1}{2}\frac{\partial
V_{p}(\tilde{\Phi})} {\partial \tilde{\Phi}} \Big ]
\Big|_{\vec{r}=0} =0 ~~~~~~~~~~~~~~~~~~~~\nonumber \\\nonumber\\
\longleftrightarrow ~~~\Big [
V_{p}(\tilde{\Phi})+\frac{1}{2}\frac{\partial
V_{p}(\tilde{\Phi})}{\partial \Phi} \Big ] \Big|_{\vec{r}=0} \cong
- \frac{1}{2} \frac{\partial F_\Phi}{\partial \Phi} \frac{\partial
V_{p}(\tilde{\Phi})}{\partial F_\Phi}\Big|_{\vec{r}=0} \,\,,
\end{eqnarray}
which shows that (4.22) is not the correct solution for $V_p
(\tilde{\Phi})$ anymore. Indeed from (5.7),
\begin{eqnarray}
V_{p} (\tilde{\Phi}) &=& e^{-2 \tilde{\Phi}} V_0 \big|_{\vec{r}=0}\nonumber \\
&=& e^{-2 \Phi} V_0 (1+ f_{\Phi}) \big|_{\vec{r}=0}\,\,,
\end{eqnarray}
where $f_{\Phi}$ is defined by
\begin{equation}
F_{\Phi} = - \frac{1}{2} \ln (1+ f_{\Phi})\,\,.
\end{equation}
So in order to admit a D$p-$brane (SM-brane) in addition to the
background NS-brane at $\vec{r}=0$, it is essential to allow
$f_{\Phi}$ to have a nonzero value at $\vec{r}=0$. That is, if we
define
\begin{equation}
f(0) \equiv V_0 e^{-\Phi} f_{\Phi} \big|_{\vec{r}=0}\,\,,
\end{equation}
(5.8) can be rewritten in the form
\begin{equation}
V_p (\tilde{\Phi}) = e^{-2\Phi} V_0 + e^{-\Phi}
\mathop{\Sigma}\limits^{\infty}_{n=0} V_n e^{n\Phi} + e^{-\Phi}
[f(0) - \mathop{\Sigma}\limits^{\infty}_{n=0} V_n  e^{n\Phi}]
\,\,,
\end{equation}
and in order that $V_p (\tilde{\Phi})$ includes the tensions of
both D-brane and NS-brane, $f(0)$ must compensate the terms for
the D-brane tension :
\begin{equation}
f(0) = \mathop{\Sigma}\limits^{\infty}_{n=0} V_n e^{n\Phi} \,\,.
\end{equation}
(5.12) is not a fine-tuning because $f(0)$ is arbitrary. Note that
$\Psi_i$ are subject to the equations of motion $\Box\Psi_i =0$
(with $\hat{g}_{ab}$ in $\Box$ replaced by $e^{4\Phi}
\hat{g}_{ab}$) which follow from (2.7). However, the boundary
values of $\Psi_i$ (or equivalently $f_{\Phi}$ at $\vec{r}=0$; see
also footnote 5) can be chosen arbitrarily so as to satisfy (5.12)
as we wish, and therefore (5.12) does not reduce to a fine-tuning.
As a result, the codimension-2 brane at $\vec{r}=0$ is now a stack
of an NS-brane and a D-brane in the presence of non-zero $\Psi_i$,
and in this case the boundary values of $\Psi_i$ are determined by
the D-brane tension.

\setcounter{equation}{0}
\section{Geometry}

Let us turn to the geometry of the ($3+p$)d spacetime. In Sec.3,
we found that ($3+p$)d metric with a codimension-2 brane at
$\vec{r}=0$ would take the form (3.30) with $e^{\Phi}$, $e^B$,
$\xi$ and $\psi$ given by (4.1), (4.9) (or (4.11)) and (4.12).
Though we have used $r-$coordinates so far, it is also suggestive
to introduce new coordinate systems.

In the case $\Lambda <0$, we introduce a new variable $\chi$
defined by
\begin{equation}
\Big( \frac{r}{r_0} \Big)^{2 |\gamma|} \,=\, \frac{(\chi
-a)}{(\chi +a)} \,\,, ~~~~~~~~(a={\rm const.})\,\,.
\end{equation}
If we take the constant $a$ as
\begin{equation}
a^2 = \frac{|m|}{2}\frac{\alpha^2_\xi}{|\Lambda| r^2_0} \,\,,
\end{equation}
then from (4.9) and (4.12), $\psi r^2$ and $i_M$ become,
respectively,
\begin{equation}
\psi r^2 \,=\, r^2_0 (\chi^2 -a^2 ) \,\,,
\end{equation}
and
\begin{equation}
i_M \,=\, d_M \Big(\frac{\chi -a}{\chi +a}\Big)^{2I_M} \,\,,
~~~~~~d_M \equiv \Big( \frac{r_0}{\tilde{r}_0} \Big)^{a_M} \,\,,
\end{equation}
where
\begin{equation}
I_M \,=\, -\frac{\epsilon}{m}\frac{a_M}{\alpha_\xi}\,\,,
~~~~~~\Big(\epsilon \equiv \frac{\gamma}{|\gamma|}\Big)\,\,.
\end{equation}
The metric (3.30) now takes the form
\begin{eqnarray}
ds^2 \,=\, d^{-(\alpha +2)}_{\Phi}
(\chi^{2}-a^{2})^{\frac{2}{m}-1-\frac{(\alpha
+2)}{2m}}\Big(\frac{\chi -a}{\chi +a}\Big)^{-(\alpha
+2)\hat{I}_\Phi} ~~~~~~~~~~~~~~~~~~~~~~~~~~~~~~\nonumber \\
\times \Big[~\frac{2}{|m \Lambda|}\frac{d\chi^2}{(\chi^2 -a^2)} +
\frac{r^2_0}{d^2_\xi}(\chi^2 -a^2)^{1-\frac{2}{m}} \Big(\frac{\chi
-a}{\chi +a}\Big)^{-2\hat{I}_\xi} d\theta^2 ~\Big]
~~~~~~~~~~~\nonumber \\+ d_B (\chi^2 -a^2)^{-\frac{(\alpha
+2)}{2m}}\Big(\frac{\chi -a}{\chi +a}\Big)^{\hat{I}_B} \big( -dt^2
+ d\vec{x}^2_p \big)\,\,,~~~~~~(\Lambda <0)\,\,,
\end{eqnarray}
where $\hat{I}_M$ is defined by
\begin{equation}
\hat{I}_M = -\frac{\epsilon}{m}\frac{\alpha_M}{\alpha_\xi} \,\,,
\end{equation}
and therefore
\begin{equation}
\hat{I}_\xi = -\frac{\epsilon}{m}\,\,,~~~~~~\hat{I}_B =
-\frac{\epsilon}{m}l \,\,,~~~~~~\hat{I}_\Phi = -\frac{\epsilon}{m}
\Big[\frac{(p+1)}{4}l - \frac{1}{2} \Big] \,\,.
\end{equation}
If $\Lambda >0$, ($\chi -a$) (and ($\chi^2 -a^2$)) in (6.6) must be
replaced by ($a- \chi$) (and ($a^2 - \chi^2$)). We further
introduce $\eta$ defined by
\begin{equation}
\chi \,=\, a \cosh \sqrt{|\Lambda|}\, \eta \,\,,
\end{equation}
then (6.6) becomes
\begin{eqnarray}
ds^2 \,=\, d^{-(\alpha +2)}_{\Phi} a^{\frac{4}{m}-2-\frac{(\alpha
+2)}{m}}\Big({\sinh \sqrt{|\Lambda|}\,
\eta}\Big)^{\frac{4}{m}-2-\frac{(\alpha +2)}{m}} \Big({\tanh
\frac{\sqrt{|\Lambda|}}{2}\, \eta}\Big)^{-2(\alpha+2)\hat{I}_\Phi} ~~~~~~~~~~~~~~~\nonumber \\
\times \Big[~\frac{2}{|m|} d\eta^2 + a^{2-\frac{4}{m}}\,
\frac{r^2_0}{d^2_\xi}
 \Big({\sinh \sqrt{|\Lambda|}\,
\eta}\Big)^{2-\frac{4}{m}}\,\Big({\tanh
\frac{\sqrt{|\Lambda|}}{2}\, \eta}\Big)^{-4\hat{I}_\xi} d\theta^2
~\Big] ~~~~~~~~~~~~~~~~~~\nonumber
\\+ d_B a^{-\frac{(\alpha +2)}{m}} \Big({\sinh \sqrt{|\Lambda|}\,
\eta}\Big)^{-\frac{(\alpha +2)}{m}} \Big({\tanh
\frac{\sqrt{|\Lambda|}}{2}\, \eta}\Big)^{2\hat{I}_B} \big(-dt^2 +
d\vec{x}^2_p \big)\,\,,~~(\Lambda <0)\,\,,~~
\end{eqnarray}
and if $\Lambda >0$, the hyperbolic functions in (6.10) should be
replaced by trigonometric functions.

For $\Lambda =0$, on the other hand, the new coordinate $\eta$ is
defined by
\begin{equation}
\frac{\eta}{\eta_0} \equiv \Big(\frac{r}{r_0} \Big)^{|\gamma|}
\,\,,
\end{equation}
and from (4.11) and (4.12) the metric (3.30) becomes
\begin{eqnarray}
ds^2 \,=\, d^{-(\alpha +2)}_{\Phi} \,
\hat{b}_{0}^{\frac{2}{m}-1-\frac{(\alpha +2)}{2m}}\,
 \Big(\frac{\eta}{\eta_0}\Big)^{2\epsilon
\, [\frac{2}{m}-1-\frac{(\alpha +2)}{2m}] -2 (\alpha
+2)\hat{I}_\Phi} ~~~~~~~~~~~~~~~~~~~~\nonumber \\
\times
\Big[~\frac{|m|}{2}\Big(\frac{\eta}{\eta_0}\Big)^{2(\epsilon -1)}
\, \frac{d\eta^2}{\eta^2_0} +
\hat{b}_0^{1-\frac{2}{m}}\,\frac{r_0^2}{d^{2}_\xi} \,
\Big(\frac{\eta}{\eta_0}\Big)^{2\epsilon (1-\frac{2}{m})-
4\hat{I}_\xi} d\theta^2 ~\Big] ~~~~~~~~~~~~~~~\nonumber
\\+ d_B \, \hat{b}_0^{-\frac{(\alpha +2)}{2m}}\, \Big(\frac{\eta}{\eta_0}\Big)^{-\epsilon \frac{(\alpha
+2)}{m} +2 \hat{I}_B} \big(-dt^2 + d\vec{x}^2_p
\big)\,\,,~~~(\Lambda=0)\,\,,
\end{eqnarray}
where we have set $b_0 /r^2_0 \equiv \hat{b}_0$. Note that if we
take
\begin{equation}
\hat{b}_0 = \frac{2}{m}|\gamma|^2 \frac{\eta_0^2}{r^2_0} \,\,,
\end{equation}
(6.10) reduces to (6.12) in the limit $\Lambda \rightarrow 0$
provided that $\sqrt{|\Lambda|}$ is replaced with $2/\eta_0$.

The above metrics become greatly simplified when $\alpha= -2$
($m=2$). For $\alpha =-2$, (6.10) becomes
\begin{equation}
ds^2 = d\eta^2 + \frac{r^2_0}{d^2_\xi} \Big({\tanh
\frac{\sqrt{|\Lambda|}}{2}\, \eta}\Big)^{-4\hat{I}_\xi} d\theta^2
+ d_B \Big({\tanh \frac{\sqrt{|\Lambda|}}{2}\,
\eta}\Big)^{2\hat{I}_B} \big(-dt^2 + d\vec{x}^2_p \big)\,\,,
~~~(\Lambda <0)\,\,,
\end{equation}
where $\hat{I}_\xi$ and $\hat{I}_B$ are given by $\hat{I}_\xi
=-\frac{1}{2}$ and $\hat{I}_B = - \frac{l}{2}$ (note that
$\epsilon(\equiv \gamma/|\gamma|)$ must be positive in the case
$\Lambda \neq 0$, see Sec.4). For $\Lambda =0$, on the other hand,
the metric is expected to be independent of $\alpha$, as is
obvious from (2.1). So in the case $\Lambda =0$ we can choose any
value for $\alpha$ (this means that (6.12) is invariant under the
change of $\alpha$), and in what follows we will always take
$\alpha= -2$ ($m=2$) in the case $\Lambda =0$. For $\alpha= -2$,
(6.12) reduces to
\begin{equation}
ds^2 = \Big(\frac{\eta}{\eta_0}\Big)^{2(\epsilon -1)} \, d\eta^2 +
\frac{r_0^2}{d^{2}_\xi} \, \Big(\frac{\eta}{\eta_0}\Big)^{-
4\hat{I}_\xi} d\theta^2 + d_B
\Big(\frac{\eta}{\eta_0}\Big)^{2\hat{I}_B} \big(-dt^2 +
d\vec{x}^2_p \big)\,\,,~~~(\Lambda =0)\,\,,
\end{equation}
where $\hat{I}_\xi =-\frac{\epsilon}{2}$ and $\hat{I}_B =
-\frac{\epsilon}{2}l$.

Finally, the dilaton is given in the $\eta-$coordinates by
\begin{equation}
e^{\Phi} ~=~
\begin{cases}
d_{\Phi} \, a^{2k_{\Phi}} \big(\sinh \sqrt{|\Lambda|}\eta
\big)^{2k_{\Phi}} \big( \tanh {\sqrt{|\Lambda|}\eta}/2
\big)^{2\hat{I}_{\Phi}(\epsilon=+1)},
~~&\text{for $\Lambda < 0$}\,\,,\\\\
d_{\Phi} \, \hat{b}_{0}^{k_{\Phi}} \big(
{\eta}/{\eta_0}\big)^{2\epsilon k_{\Phi} +2\hat{I}_{\Phi}},
~~&\text{for $\Lambda =0$ ,}
\end{cases}
\end{equation}
which reduces for $\alpha= -2$ ($m=2$) and $l=0$ to
\begin{equation}
e^{\Phi} ~=~
\begin{cases}
g_s /\cosh \sqrt{|\Lambda|}\eta/2  \,\,,
~~&\text{for $\Lambda < 0$}\,\,,\\\\
g_s \,\,, ~~&\text{for $\Lambda =0$ ,}
\end{cases}
\end{equation}
where $g_s$ is defined by
\begin{equation}
g_s ~=~
\begin{cases}
d_{\Phi} \, a^{-\frac{1}{2}}/\sqrt{2}\,\,\,\,, ~~~{\rm for~\Lambda<0 }\,\,,\\\\
d_{\Phi} \, \hat{b}_{0}^{-\frac{1}{4}} \,\,\,\,, ~~~{\rm
for~\Lambda=0 } \,\,.
\end{cases}
\end{equation}
The constant $g_s$ in (6.17) plays the same role as the asymptotic
value $g_s$ of the theory with an ordinary NS5-brane. It can be
taken to have any arbitrary desired value in (6.18) by choosing
$d_{\Phi}$ properly. In our case we will take $g_s \rightarrow 0$,
which corresponds to $d_{\Phi} \rightarrow 0$. The theory with
NS-branes (or D-branes) in the limit $g_s \rightarrow 0$ is not
new as mentioned in introduction. Such an idea can be found in the
literatures on LST, where they consider the limit $g_s \rightarrow
0$ by which the bulk degrees of freedom decouple from the degrees
of freedom of the brane, and one is left with physics on the
brane. (Another example of using $g_s \rightarrow 0$ can be found
in AdS/CFT correspondence where the gauge coupling $g^2_{\mathrm
YM}(=g_{s})$ goes to zero, while the rank $N$ of the gauge group
(or the number of D-branes) goes to infinity in such a way that
$Ng^2_{\mathrm YM}$ is held fixed.) In this paper we are
essentially considering the same limit as those of these theories.
Also, there is another important reason for considering the limit
$g_s \rightarrow 0$. In our brane world models, taking $g_s
\rightarrow 0$ naturally accords with the hierarchy conjecture as
we will see in Sec.11. If we assume $M_s \sim TeV$ and $\rho_{max}
\sim TeV^{-1}$ (where $M_s$ is the string scale: $M_s =
1/\sqrt{\alpha^{\prime}}$, and $\rho_{max}$ is the size of
$\Sigma_2$), $g_s$ is estimated to be $g_s \sim 10^{-16}$, which,
however, is just a realistic value of the decoupling limit $g_s
\rightarrow 0$.

(6.17) shows that $e^{\Phi}$ becomes the constant $g_s$ as we
approach the background NS-brane ($\eta \rightarrow 0$). This
contrasts with the case of the ordinary NS5-brane of the type IIA
or type IIB string theory where $e^{\Phi}$ diverges in the
vicinity of the NS5-brane. The reason why this happens is that the
NS-brane discussed in the present paper is not the real BPS object
of the string theory (see footnote 2). Had it been a BPS-brane
with (magnetic) charge $N$, we would have had $e^{2\Phi} \sim
-g^2_s N \ln \eta$ (for $\Lambda =0$) near $\eta =0$, as can be
inferred from the usual NS5-brane solution $e^{2\Phi} = g^2_s
(1+N\alpha^{\prime}/\eta^2)$ which becomes $e^{2\Phi} \sim g^2_s N
\alpha^{\prime}/\eta^2$ as $\eta \rightarrow 0$. So $e^{2\Phi}$
would have diverged as $\eta \rightarrow 0$, just as in the case
of the usual NS5-brane. In general $e^{\Phi}$ diverges in the
vicinity of the (BPS) NS-branes, and this divergence near
singularity gives rise to an introduction of a new parameter
($\equiv g_{lst}$) in LST \cite{20,21}, which serves as an
effective string coupling on the D-branes stretched between $N$
background NS5-branes. We will be back to this point later (see
footnote 4).

By (6.18), one can estimate the constant $r_0^2 /d^2_\xi$
appearing in the metrics (6.14) and (6.15). Using (6.2), (6.13),
(6.18), and the relation $d_{\xi}=d^{-2}_{\Phi} d^2_B$ which
follows from $i_{\xi}=i^{-2}_{\Phi} i^2_B$, one finds for $l=0$
(and for $\alpha=-2$)
\begin{equation}
\frac{r_0^2}{d_{\xi}^2} ~=~
\begin{cases}
4 g^4_s \alpha^2_{\xi} / |\Lambda|\,\,, ~~&\text{for $\Lambda < 0$}\,\,,\\\\
g^4_s \alpha^2_{\xi} \eta^2_0 \,\,, ~~&\text{for $\Lambda =
0$}\,\,,
\end{cases}
\end{equation}
where we have used the fact that $d_B=(r_0
/\tilde{r}_{0})^{l\alpha_{\xi}}$ when $\alpha =-2$ (see (6.4) and
(4.18)), so $d_B = 1$ for $l=0$. After all this, we observe that
the $(3+p)$d metrics essentially depend on $V_p (\Phi)$ only
through the constants $l$ and $\alpha_{\xi}$. In the next section
we will show that the shifts in $l$ and $\alpha_{\xi}$ due to
quantum corrections to the D-brane tension are of an order $g^2_s$
, which implies that the change of the bulk geometry (including
flat intrinsic geometry of the brane) due to quantum corrections
to the D-brane tension is extremely suppressed in the limit $g_s
\rightarrow 0$.

\setcounter{equation}{0}
\section{$\bf {V_{p}(\Phi)}$-independent bulk geometry}

In the usual self-tuning brane world models, the intrinsic
geometry of the brane is not affected by the brane tension
$V_{p}(\Phi)$, but the geometry of bulk spacetime is always
affected by (the change of) $V_{p}(\Phi)$. For instance in 6d (or
codimension-2 brane world) models, the presence of a flat brane
with the tension $V_{p}(\Phi)$ causes a deficit angle in the
transverse dimensions the magnitude of which is proportional to
$V_{p}(\Phi)$. Thus a change $\delta V_{p}(\Phi)$ in $V_p (\Phi)$
necessarily causes a corresponding change in the deficit angle,
and this could lead to a failure of the self-tuning scheme. The
$V_{p}(\Phi)-$dependency of the bulk geometry, however, can be
avoided by introducing background NS-brane; it is extremely
suppressed in the limit $g_s \rightarrow 0$ due to the presence of
the NS-brane. In this section we will show that the bulk geometry
is really practically unaffected by the shift $\delta V_{p}(\Phi)$
in the limit $g_s \rightarrow 0$.

In Sec.5 we have seen that the solutions with $\Psi_i \neq 0$
(i.e., $\hat{l} \neq 0$) are basically self-tuning solutions. With
$\hat{l} \neq 0$, the relation (4.18) can be expressed in terms of
$V_p (\Phi)$ and $\partial V_p (\Phi)/\partial \Phi$ as
\begin{equation}
\alpha V_{p}(\Phi) + \frac{(\alpha -2)}{4}\, \frac{\partial
V_{p}(\Phi)}{\partial \Phi} \,=\, l \, \Big[ \frac{(\alpha
+2)(p+1)}{4} V_{p}(\Phi) + \frac{[(\alpha +2)(p+1)+4]}{8}\,
\frac{\partial V_{p}(\Phi)}{\partial \Phi} \,\Big]\,\,,
\end{equation}
where $V_{p}(\Phi)$ represents a sum of the tensions of the
D-brane and the NS-brane; $V_{p}(\Phi) \equiv V_D (\Phi)+ V_{NS}
(\Phi)$, each of which is assumed to take the form
\begin{equation}
V_{D} (\Phi) = e^{-\Phi} \sum_{n=0}^{\infty} V^{(D)}_{n} e^{n
\Phi}\,\,, ~~~~~ V_{NS} (\Phi) = e^{-2\Phi} V_0^{(NS)} \,\,.
\end{equation}
Substituting (7.2) into (7.1) gives an $\infty-$th order equation
for $g_s$($\equiv e^{\Phi}\big|_{\eta=0}$):
\begin{equation}
\sum_{n=0}^{\infty} c^{(l)}_{n} g^{n}_{s} =0 \,\,,
\end{equation}
where
\begin{equation}
c^{(l)}_{n} =\big[ k_1 + (n-1) k_2  \big] V^{(D)}_{n} g_{s} +
\big[k_1 + (n-2) k_2  \big] V^{(NS)}_{0} \delta_{n0}\,\,,\nonumber
\end{equation}

\begin{equation}
k_1 \equiv - \alpha  + \frac{(\alpha +2)(p+1)}{4}l  \,\,, ~~~ k_2
\equiv \frac{(l+2)}{2} + \frac{(\alpha +2)}{2} \Big[
\frac{(p+1)}{4}l -\frac{1}{2}\Big]\,\,.
\end{equation}
Since $\alpha_M$ are combinations of $V_{p}(\Phi)$ and
$dV_{p}(\Phi)/d\Phi$, the constant $l$ would also depend on
$V_{p}(\Phi)$ and $dV_{p}(\Phi)/d\Phi$, or equivalently on
infinite numbers of $V_{n}^{(D)}$ and $V_{0}^{(NS)}$. Thus solving
(7.3) for $l$ (assuming that it can be solved) gives $l$ in terms
of $V_n^{(D)}$ and $V_{0}^{(NS)}$ for a given value of $g_s$.

Though we wish to solve (7.3) for $l$ for a given value of $g_s$,
it is convenient to begin with an assumption that (7.3) is an
$n-$th order ($\infty-$th order in fact) equation for $g_s$. To
solve (7.3), rewrite it as
\begin{eqnarray}
-\Bigg\{l + \frac{(\alpha +2)}{2} \Bigg\}
V^{(NS)}_{0}\,+\,\Bigg\{\big(1-\frac{l}{2}\big)+(\alpha
+2)\Big[\,\frac{(p+1)}{8}l-\frac{3}{4}\,\Big]\Bigg\}V^{(D)}_{0}\,g_s
\nonumber \\ +\,\Bigg\{2  + (\alpha +2)\Big[\frac{(p+1)}{4}l-1
\,\Big] \Bigg\} V^{(D)}_{1}\,g_s^2 \,+\,\cdots \,=\,
0\,\,.~~~~~~~~~~
\end{eqnarray}
(7.5) has a peculiar form. The $n$-th order coefficients
$V_n^{(D)}$ appear as an $(n+1)$-th order coefficients in (7.5).
Equation (7.5) admits real solutions representing the limit $g_s
\rightarrow 0$ for a certain value of $l$. Neglecting higher-order
terms for a moment (and assuming that $l$ is not infinitely
large), one finds that the first two terms in (7.5) can cancel
each other when $\hat{l}\,V_0^{(NS)}$ is much smaller than
$V_0^{(D)}$. To be precise, (7.5) requires
\begin{equation}
\hat{l}\,\sim \, \Big(V_0^{(D)}/V_0^{(NS)} \Big) g_s \,\,,
\end{equation}
provided the higher order terms are neglected. This is important.
(7.6) implies that $\hat{l}$ should be as small as $g_s$ if
$V_0^{(NS)}$ is of the same order as $V_0^{(D)}$ ; i.e., $\hat{l}
\sim g_s$ if $V_0^{(NS)} \sim V_0^{(D)}$. (In string theory,
$V_0^{(NS)}$ and $V_0^{(D)}$ both take the same value
$V_0^{(D,NS)} \sim 1/{\alpha^{\prime}}^3$ for $p=5$.) So if $g_s
\rightarrow 0$, $\hat{I}_B$ and similarly $\hat{I}_{\Phi}$ are
practically not different from their values with $\hat{l}=0$. In
the previous section we have observed that the $(3+p)$d metrics
depend on $V_{p}(\Phi)$ only through the constants $l$ (i.e.,
$\hat{l}$) and $\alpha_\xi$. Apart from $\alpha_\xi$, this implies
that the bulk geometry is practically unchanged by an addition of
the SM-brane (the D-brane) to the background NS-brane. Note that
$\Sigma_2$ with an NS-brane alone corresponds to $\hat{l}=0$
($\Psi_i =0$), while $\Sigma_2$ with both SM- and NS-branes
corresponds to $\hat{l} \neq 0$ ($\Psi_i \neq 0$).

Once $\hat{l}$ is determined at the tree-level for the given
values of $V_0^{(D,NS)}$, the effect of the higher order terms can
be obtained by adding $\delta l$ to $\hat{l}$, where $\delta l$($=
\delta \hat{l}$) is the shift in $l$ due to quantum corrections to
the tension of the SM-brane. As is obvious from (7.5), $\delta l$
is proportional to $g_s^2$; i.e.,
\begin{equation}
\delta l \sim \Big( V_1^{(D)} /V_0^{(NS)} \Big) g^2_s \,\,.
\end{equation}
To estimate the magnitude of $\delta l$, first consider the case
where the gauge coupling is simply given by $g^2_{\mathrm YM} \sim
g_s {\alpha^{\prime}}^{(p-3)/2}$, and $V_1^{(D)} /V_0^{(NS)}$ is
of order the unity. In this case $\delta l$ is simply
\begin{equation}
\delta l \sim g^4_{\mathrm YM}\, {\alpha^{\prime}}^{-(p-3)}\,\,,
\end{equation}
i.e., the shift in $l$ due to quantum corrections to the brane
tension is suppressed with the factor $g^4_{\mathrm YM}$.

There is a different way of viewing (7.7). Suppose that $V_D
(\Phi)$ is written in the form
\begin{equation}
V_D (\Phi) = \frac{V_0^{(D)}}{g_s}\, \Big(1+
\mathop{\Sigma}\limits^{\infty}_{n=1} m_n g^n \Big) \,\,,
\end{equation}
where the terms with $n\geq 1$ describe the quantum corrections to
the brane tension due to SM-fields living on the D-brane. The
coefficients $m_n$ are dimensionless, and $m_n \sim \large{O}(1)$
provided $n$ is not very large. The constant $g$ is a
dimensionless (effective) coupling defined on the D-brane, and
$g_{\mathrm YM}$ is now given by $g_{\mathrm YM}^2 \sim g
{\alpha^{\prime}}^{(p-3)/2}$, while $g_s$ is taken to be $g_s
\rightarrow 0$. The expansion (7.9) may be applied to both cases
where the quantum corrections are dilaton dependent, and where the
quantum corrections are dilaton independent. In the former case
the coupling $g$ is given in terms of $g_s$ (see (7.10)), and it
becomes an analogue of $g_{lst}$ of LST\footnote{As mentioned in
Sec.6, LST admits an effective coupling $g_{lst}$ on the D-brane
\cite{20}. $g_{lst}$ is defined by $g_{lst} \sim g_s/LM_s$, where
$L$ represents the separation of the $N$ NS-branes which are
distributed uniformly on a transverse circle in moduli space. In
the double scaling limit $g_s$,$L \rightarrow 0$ with $g_s /L$
held fixed, $g_{lst}$ takes a certain finite value and it plays a
role of the effective string coupling on the D-branes stretched
between $N$ NS-branes. So in order to define $g_{lst}$, it is
necessary to have a configuration that there are $N$ (BPS)
NS-branes sitting around the singularity. In the present paper,
however, we are only considering a brane world scenario which uses
LST, rather than being LST itself, only as a partial analogue of
the theory, and we simply assume (without extending to the
configuration with $N$ (BPS) NS-branes) that an effective coupling
$g$ (which is dimensionless, and associated with $g_{\mathrm YM}$
by the equation $g_{\mathrm YM}^2 \sim g
{\alpha^{\prime}}^{(p-3)/2}$) is intrinsically defined on the
D-brane (SM-brane) as an analogue of $g_{lst}$ of LST.} where the
effective coupling on the D-brane is given by $g_{lst}$, while the
coupling to the bulk modes behaves as $g_s$ (i.e., while $V_D
(\Phi)$ is given by $V_D (\Phi) \sim 1/
{\alpha^{\prime}}^{(p+1)/2} g_s$ at the tree level, its quantum
corrections should be expanded in the SM coupling $g_{\mathrm
YM}$, or equivalently $g_{lst}$). Comparing (7.9) with (7.2) one
finds that
\begin{equation}
g = \Big( \frac{1}{m_n} \frac{V_n^{(D)}}{V_0^{(D)}}\Big)^{1/n}\,
g_s \,\,,
\end{equation}
and (7.7) becomes
\begin{equation}
\delta l \sim g^2_{\mathrm YM}\, g_s \,
{\alpha^{\prime}}^{-(p-3)/2}\,\,.
\end{equation}
So the shift $\delta l$ is suppressed with the factor $\sim
g^2_{\mathrm YM}\, g_s$ this time. Finally, in the case where the
quantum corrections are dilaton independent, $g$ simply represents
the SM coupling $g_{\mathrm YM}$ (through the equation $g_{\mathrm
YM}^2 \sim g {\alpha^{\prime}}^{(p-3)/2}$) which is now
independent of $g_s$. But in this case too, one can show that
$\delta l$ is also given by (7.11). So in any case, the change of
the bulk geometry due to quantum corrections is extremely
suppressed in the limit $g_s \rightarrow 0$.

In the above discussion we have implicitly assumed that (7.3) can
be solved for $g_s$. But in reality, it is impossible to solve the
$\infty-$th order equation, and we are only allowed to solve the
$n$-th order equation with finite $n$ (perhaps for $n \leq 4$).
For instance if $n=2$, the equation takes the form
\begin{equation}
\hat{c}^{(l)}_2 g^2_s  + \hat{c}^{(l)}_1 g_s + \hat{c}^{(l)}_0
=0\,\,,
\end{equation}
where $\hat{c}_n^{(l)}$ do not include $g_s$. (7.12) admits real
solutions for $g_s$ as long as the condition
\begin{equation}
(\hat{c}^{(l)}_1 )^2 - 4 \hat{c}^{(l)}_2 \hat{c}^{(l)}_0 \geq 0
\end{equation}
is met. For (7.5), and for $\alpha =-2$ for simplicity, the
condition (7.13) reads
\begin{equation}
(1-\frac{l}{2})^2 \,(V^{(D)}_{0})^2 \,+\, 8l V^{(NS)}_{0}
V^{(D)}_{1} \geq 0 \,\,,
\end{equation}
and the solutions becomes
\begin{equation}
g_s = \bigg[-(1-\frac{l}{2}) V^{(D)}_{0} \pm
\sqrt{(1-\frac{l}{2})^2 (V^{(D)}_{0})^2 \,+\, 8l \, V^{(NS)}_{0}
V^{(D)}_{1}} \,\, \bigg] / 4 V^{(D)}_{1} \,\,.
\end{equation}
(7.14) does not lead to a fine-tuning of $V^{(NS,D)}_{n}$. It only
restricts the ranges of $V^{(NS,D)}_{n}$. Also one can check that
one of the solutions in (7.15) reduces to $g_s \sim l
V^{(NS)}_{0}/ V^{(D)}_{0}$ for $l \rightarrow 0$, which is just
the one that we have obtained in (7.6). The other solution in
(7.15) reduces (upon using $V_0^{(D)} \sim V_0^{(NS)}$) to $g_s
\sim - V^{(D)}_{0}/ 2 V^{(D)}_{1}$ for $l \rightarrow 0$, which
may correspond to the strongly coupled case for either
$V^{(D)}_{0}/ V^{(D)}_{1} \sim \large{O}(1)$ or $V^{(D)}_{0}/
V^{(D)}_{1} \sim g_s/g$ (see (7.10)), and should perhaps be
discarded in the framework using perturbation.

So far we have concentrated on the constant $l$. But the $(3+p)$d
metrics also depend on $\alpha_\xi$ (as well as $l$) as observed
in Sec.6. But still, one can show that their geometries are
virtually unchanged by an addition of the SM-brane though the
effect of $\delta \alpha_{\xi}$ is taken into account. In order to
see this, define a constant $k$ as
\begin{equation}
k = \frac{\alpha_\xi (V_D (\Phi) + V_{NS} (\Phi)) - \alpha_{\xi}
(V_{NS} (\Phi))}{\alpha_{\xi} (V_{NS} (\Phi))} \, = \,
\frac{\delta \alpha_\xi}{\alpha_{\xi} (V_{NS} (\Phi))}\,\,,
\end{equation}
where $\alpha_\xi (V_D (\Phi) + V_{NS} (\Phi))$ represents the
value of $\alpha_\xi$ when the brane at $\eta=0$ is a coincident
brane of the SM- and the background NS-brane, while $\alpha_{\xi}
(V_{NS} (\Phi))$ the value of $\alpha_\xi$ when the brane at
$\eta=0$ is simply an NS-brane. From (4.7), (7.2) and (7.16), one
obtains
\begin{equation}
k V_0^{(NS)} + \frac{1}{2} \Big[\frac{(\alpha +2)(p+1)}{4} -1
\Big] V_0^{(D)}g_s + \frac{(\alpha +2)(p+1)}{4} V_1^{(D)} g_s^2 +
\cdots =0 \,\,,
\end{equation}
and the same analysis made for (7.5) can be applied just as it is
to this case too. One finds $k \sim g_s $, implying (together with
$\hat{l} \sim g_s$) that the $(3+p)$d geometries are virtually
unaffected by an addition of the SM-brane in the limit $g_s
\rightarrow 0$. Similarly, the shift in $k$ becomes $\delta k \sim
g^2_s$ (In the case $\alpha =-2$, it is even smaller; i.e.,
$\delta k \sim g^3_s$.) as before, so the change of the $(3+p)$d
geometry due to quantum corrections is extremely suppressed in the
limit $g_s \rightarrow 0$.

\setcounter{equation}{0}
\section{Codimension-1 brane as a T-dual of codimension-2 brane}

In Sec.4, we found that solution to the set of field equations
takes the form of (4.1) with $\psi (r)$ given by (4.9) or (4.11)
according to whether $\Lambda \neq 0$ or $\Lambda =0$. In the case
$\Lambda \neq 0$, (4.9) is valid only when $\gamma$ is positive.
But in the case $\Lambda =0$, (4.11) is valid for both positive
and negative $\gamma$. So far we have assumed that the brane at
$r=0$ (or $\eta =0$) is a codimension-2 brane. But codimension-2
brane can exist only when $\Sigma_2$ is closed at $r=0$. In some
cases $\Sigma_2$ fails to be closed at $r=0$; rather, it spreads
out (i.e., $\sqrt{g_{\theta\theta}}$ diverges) there. Whether
$\Sigma_2$ is closed at $r=0$ or not entirely depends on the
signature of $\gamma$. When $\Lambda \neq 0$, $\Sigma_2$ is always
closed at $r=0$ ($\eta =0$) because $\gamma$ must be positive in
the case $\Lambda \neq 0$, and $2- (4/m) -4 \hat{I}_\xi =2
>0$ in (6.10) for positive $\gamma$. When $\Lambda =0$, on the other hand,
(6.15) shows that while $\Sigma_2$ is closed at $r=0$ ($\eta=0$)
if $\epsilon > 0$ ($r>0$), it spreads out as $r \rightarrow 0$
($\eta \rightarrow 0$) if $\epsilon < 0$ ($r<0$). So the brane at
$r=0$ is a codimension-2 brane in the case $\epsilon > 0$, while
it is necessarily a codimension-1 brane in the case $\epsilon <
0$. In the following discussion, we will identify these branes as
certain limits of the type II codimension-1 brane introduced in
Appendix.

Once these branes (of the case $\Lambda=0$) are identified with
the type II codimension-1 branes, we observe that they are T-duals
of each other provided that they have the same mass. In order to
see this, return to the metric (6.15). For $\epsilon > 0$, (6.15)
becomes
\begin{equation}
ds^2_{\epsilon > 0} = d\eta^2 +
\frac{r^2_0}{d^2_{\xi}}\big(\frac{\eta}{\eta_0} \big)^2 d\theta^2
+ d_B \big(\frac{\eta_0}{\eta} \big)^{l} \big(-dt^2 + d\vec{x}^2_p
\big) \,\,,
\end{equation}
while for $\epsilon < 0$,
\begin{equation}
ds^2_{\epsilon < 0} = \big(\frac{\eta_0}{\eta} \big)^{4} d\eta^2 +
\frac{r^2_0}{d^2_{\xi}}\big(\frac{\eta_0}{\eta} \big)^2 d\theta^2
+ d_B \big(\frac{\eta}{\eta_0} \big)^{l} \big(-dt^2 + d\vec{x}^2_p
\big) \,\,.
\end{equation}
These two metrics are related to each other by a duality
transformation. We see that one of them is converted into another
by a transformation
\begin{equation}
\eta \rightarrow \tilde{\eta}=\frac{\eta^2_0}{\eta}\,\,,
\end{equation}
and if we identify $\eta_0$ with the string length
$\sqrt{\alpha^{\prime}}$, (8.3) becomes a (closed string)
T-duality transformation of the string theory. Codimension-1
branes can be obtained from (8.1) and (8.2) by fixing $\eta$ to a
constant ; i.e., $\eta=\eta_c$, where $\eta_c$ represents the
position (or the radius) of the branes in the $\eta-$coordinates.
If these two codimension-1 branes have the same mass, they are
T-duals of each other because they are related by the duality
transformation (8.3), and the total mass of the brane is conserved
under duality transformation. Let us identify these branes as the
type II codimension-1 branes, and take $\eta_c \rightarrow 0$. The
codimension-1 brane of the case $\epsilon > 0$ then shrinks by
$\eta_c \rightarrow 0$ to a point to become a codimension-2 brane,
while the other one of the case $\epsilon < 0$ still remains to be
a codimension-1 brane. But still, these two branes are related by
(8.3), and they have the same mass because the total mass of the
type II codimension-1 brane is preserved under the variation of
$\eta_c$. So they are T-duals of each other.

Let us consider the codimension-1 brane of the case $\epsilon <
0$. We assume that this codimension-1 brane is located at $r=0$
for the moment. Since this brane exists only when $\Lambda =0$, it
is described by
\begin{equation}
\psi r^2 = \frac{b_0}{\big(|r|/{r_0}\big)^{-2\gamma}} \,\,,
\end{equation}
and
\begin{equation}
i_M (r)= d_M \big(\frac{|r|}{r_0} \big)^{a_M} \,\,.
\end{equation}
One can check that (8.4) and (8.5) satisfy the field equations
(A.13) with $\Lambda =0$, and (A.14), respectively. (8.4) and
(8.5) show that there is a reflection symmetry about the brane at
$r=0$. We take this codimension-1 brane as an orbifold fixed line;
i.e., we identify every point of the region $r<0$ with the
corresponding point of the region $r>0$, and then take the region
$r>0$ as a fundamental domain. Note that extra factor two has been
multiplied on each $\alpha_M$ in (A.13) and (A.14). It reflects
the fact that the orbifold fixed line at $r=0$ is an overlap of
two codimension-1 branes each of which belongs to the
corresponding regions $r<0$ and $r>0$. So the tension $V_{p+1}
(\Phi)$ (and consequently $\alpha_M$) must be doubled.

So far we have assumed that the type II codimension-1 brane is
placed at $r=0$. However, we want the brane at $r=0$ to be a
codimension-2 brane because we want to set our SM-brane (a
codimension-2 brane) at $\vec{r}=0$. Thus in the followings, the
type II codimension-1 brane will be moved to $r=r_m$ (or $\eta
=\eta_m$ in the $\eta-$coordinates), and it will serve as a T-dual
of the codimension-2 brane at $\vec{r}=0$ (see the case III of
Sec.10). So in the present paper we always consider the case where
$\Sigma_2$ is closed at $r=0$.

\setcounter{equation}{0}
\section{Matching conditions}

 According to the metrics in Sec.6, two extra dimensions of $\Sigma_2$ form an
infinite volume space, which may need to be compactified anyhow.
In order to avoid such noncompact extra dimensions, we introduce a
codimension-1 brane ($\equiv$ brane B) at a finite distance from
the codimension-2 brane at $\vec{r}=0$. Because this codimension-1
brane is expected to be an ordinary codimension-1 brane, we begin
with the field equations for the type I codimension-1 brane
introduced in Appendix. We assume that the brane B is placed at
$r=r_B$.

The field equations (A.3)-(A.5) require the fields to satisfy
the matching conditions
\begin{equation} r \frac{d\ln R_{II}}{dr}
\Big|_{r=r_B} - r \frac{d\ln R_{I}}{dr} \Big|_{r=r_B} = - \kappa^2
C_1^{(p+1)} \,\,,
\end{equation}
\begin{equation}
r \frac{d\Phi_{II}}{dr} \Big|_{r=r_B} - r \frac{d\Phi_{I}}{dr}
\Big|_{r=r_B} = \frac{(p+1)}{2}\kappa^2 C_2^{(p+1)} + \frac{1}{2}
\kappa^2 C_1^{(p+1)} \,\,,
\end{equation}
\begin{equation}
r \frac{dB_{II}}{dr} \Big|_{r=r_B} - r \frac{dB_{I}}{dr}
\Big|_{r=r_B} = 2\kappa^2 C_2^{(p+1)} \,\,,
\end{equation}
in addition to
\begin{equation}
R_{I}\big|_{r=r_B} = R_{II}\big|_{r=r_B} \,\,, ~~~~~e^{\Phi_{I}}
\big|_{r=r_B} = e^{\Phi_{II}} \big|_{r=r_B} \,\,, ~~~~~e^{B_{I}}
\big|_{r=r_B} = e^{B_{II}} \big|_{r=r_B} \,\,,
\end{equation}
where the indices I and II represent the regions $r<r_B$ and
$r>r_B$, respectively. In the region I, a coincident brane (of the
SM- and the background NS-branes) is placed at $\vec{r}=0$, and we
want this to be a codimension-2 brane (see Sec.8). In the region
II, we want a corresponding brane to be placed at $r=2\,r_B
(\equiv r_{m})$ by reason of symmetry. So the configuration is
that we have a codimension-2 brane at $\vec{r}=0$, and a
corresponding brane at $r=r_m$ (which can be either the type II
codimension-1 brane, or a codimension-2 brane), and finally a type
I codimension-1 brane ($=$ brane B) in the middle, i.e., at
$r=r_B$. The whole internal space $\Sigma_2$ thus consists of two
parts; i.e., the region I($\equiv 0<r<r_B \equiv \Sigma_{2I}$) and
the region II($\equiv r_B<r<r_m \equiv \Sigma_{2II}$), where $r_m
= 2 r_B$ as defined above.

In the region I, $\psi(r)$ and $i_M (r)$ are due to the brane at
$\vec{r}=0$ and directly given by (4.9)(or (4.11)) and (4.12),
respectively. In the region II, $\psi(r)$ and $i_M (r)$ are due to
the brane at $r=r_m$ and also given by (4.9)(or (4.11)) and
(4.12), but this time $r$ is replaced with $r_m -r$. That is, we
construct $\Sigma_2$ by gluing $\Sigma_{2II}$ onto $\Sigma_{2I}$
with left and right reversed. We have
\begin{equation}
\psi (r) =
\begin{cases}
{b_{0I}}\,r^{-2} \Big[ c_{1I} \big({r}/{r_{0I}}\big)^{\gamma_I}
\pm c_{2I} \big({r}/{r_{0I}}\big)^{-\gamma_I} \Big]^{-2}
&\text{(region I)
}\\\\
{b_{0II}}\,r^{-2} \Big[ c_{1II} \big( {(r_m -r)
}/{r_{0II}}\big)^{\gamma_{II}} \pm c_{2II} \big( {(r_m -r)
}/{r_{0II}}\big)^{-\gamma_{II}} \Big]^{-2} &\text{(region II),}
\end{cases}
\end{equation}
and
\begin{equation}
i_M (r) ~=~
\begin{cases}
\big( r/\tilde{r}_{0I} \big)^{\alpha_{MI}}
~~~&\text{(region I)}\\\\
\big((r_m -r)/\tilde{r}_{0II} \big)^{\alpha_{MII}}
~~~~~&\text{(region II),}
\end{cases}~~~~~~~~~~~~~~~~~~~~~~~~~~~~~~
\end{equation}
where $c_i$ are given by $c_1 =c_2 =1$ for $\Lambda \neq 0$, and
$c_1 =0$, $c_2 =1$ for $\Lambda =0$. Also, the constant $b_0$ is
$b_0 = \pm (8 \gamma^2 /m \Lambda)$ for $\Lambda \neq 0$, and
arbitrary for $\Lambda =0$. Substituting (9.5) and (9.6) into
(9.1)-(9.3) gives
\begin{equation}
\alpha_{RI} + \alpha_{RII} = \mathcal{B}_1 \,\,, ~~~ \alpha_{\Phi
I} + \alpha_{\Phi II} = -\frac{(p+1)}{2}\mathcal{B}_2
-\frac{1}{2}\mathcal{B}_1 \,\,,~~~ \alpha_{BI} + \alpha_{BII} =
-2\mathcal{B}_2 \,\,,
\end{equation}
where $\mathcal{B}_1$ and $\mathcal{B}_2$ are given by
\begin{equation}
\mathcal{B}_1 = \kappa^2 C^{(p+1)}_{1} + \frac{2}{m} Y_0 \,\,,
~~~~~ \mathcal{B}_2 = \kappa^2 C^{(p+1)}_{2} + \frac{(\alpha
+2)}{2m} Y_0 \,\,,
\end{equation}
where
\begin{equation}
Y_0 \equiv \gamma_I \frac{[c_{1I} X_I \mp c_{2I} X_I^{-1}]}{[c_{1I} X_I
\pm c_{2I} X_I^{-1}]} + \gamma_{II} \frac{[c_{1II} X_{II} \mp
c_{2II} X_{II}^{-1}]}{[c_{1II} X_{II} \pm c_{2II}
X_{II}^{-1}]}\,\,, ~~~~~\Big(\, X_{I, II} \equiv \big(
\frac{r_B}{r_{0I,II}}\big)^{\gamma_{I, II}}\Big)\,\,.
\end{equation}
Recalling (4.16), one finds that (9.7) can be solved by
setting
\begin{equation}
\mathcal{B}_1 = \mathcal{B}_2 =0 \,\,,
\end{equation}
and therefore
\begin{equation}
\alpha_{MI} + \alpha_{MII} =0 \,\,,~~~~~(M \equiv R,\,\, \Phi
,\,\,B)\,\,.
\end{equation}

With $\Psi_i =0$, (9.11) reduces to a fine-tuning of $V_p (\Phi)$.
But in the case $\Lambda=0$, once $\Psi_i$ is "turned on" (9.11)
does not restrict $V_p (\Phi)$ anymore. The reason is as follows.
In (9.11), the number of independent equations is only two
(instead of three) because they are related with each other by
(4.16). The number of independent $\alpha_M$ is also two because
they depend only on two independent functions $V_p (\Phi)$ and
$\partial V_p (\Phi)/\partial \Phi$ through $C_1$ and $C_2$; i.e.,
four $\alpha_M$ contain only two independent degrees of freedom
(indeed, four $\alpha_M$ with $M \equiv R,\,\,\Phi,\,\,B,\,\,\xi$
are related with each other by (4.15), and (4.16) as well). So the
two degrees of freedom $V_p (\Phi)$ and $\partial V_p
(\Phi)/\partial \Phi$ are fixed by the equations in (9.11). If
$\Psi_i$ is "turned on", however, the situation changes. $C_1$ and
$C_2$ (and therefore $\alpha_M$) are now going to include two more
degrees of freedom $F_{\Phi}$ and $F_{B}$ (see (3.26) and (5.2));
i.e., $\alpha_M$ become $\alpha_M =\alpha_M (V_p (\Phi),\,\,
\partial V_p (\Phi)/\partial \Phi ,\,\, F_{\Phi},\,\,F_{B})$. Thus
this time the two independent equations in (9.11) can not restrict
$V_p (\Phi)$ anymore\footnote{In this case, (9.11) only determines
the boundary values of $F_{\Phi}$ and $F_B$, or equivalently the
values of $\Psi_i$ at $r=0$ and $r=r_m$. $\Psi_i$ have been
introduced in the action in order to compensate the D-brane
tensions at the boundaries $r=0$ and $r=r_m$, where the D-branes
are assumed to be located. So, what really matters is just the
boundary values of $\Psi_i$ at $r=0$ and $r=r_m$, not the
functional form of $\Psi_i$ between $r=0$ and $r=r_m$. Indeed, the
equations of motion for $\Psi_i$ are second order equations, i.e.,
$\Box \Psi_{i}=0$ with $\hat{g}_{ab}$ in $\Box$ replaced by
$e^{4\Phi}\hat{g}_{ab}$ as can be obtained from (2.7). So the
boundary values of $\Psi_i$ at both $r=0$ and $r=r_m$ can be
chosen to compensate the D-brane tensions as we wish.} due to the
presence of the extra degrees of freedom $F_{\Phi}$ and $F_{B}$,
and therefore (9.11) does not reduce to a fine-tuning of $V_p
(\Phi)$ when $\Psi_i \neq 0$. In the case $\Lambda \neq 0$,
however, the self-tuning of $V_p (\Phi)$ is not guaranteed since
in this case we have only one extra degree of freedom, i.e., $F$.

Turning back to the matching conditions, it is convenient to replace the conditions in (9.4) by
\begin{equation}
\Psi_{I} \big|_{r=r_B} \,=\, \Psi_{II} \big|_{r=r_B} \,\,,
\end{equation}
and
\begin{equation}
i_{MI} \big|_{r=r_B} \,=\, i_{MII} \big|_{r=r_B} \,\,.
\end{equation}
The condition (9.13) can be easily satisfied if we take
\begin{equation}
\tilde{r}_{0I} = \tilde{r}_{0II} \,=\, r_B \,\,.
\end{equation}
But (9.12) gives a condition
\begin{equation}
{b_{0I}}\Big[ c_{1I} X_{I} \pm c_{2I} X^{-1}_{I} \Big] \,=\,
{b_{0II}}\Big[ c_{1II} X_{II} \pm c_{2II} X^{-1}_{II} \Big]\,\,,
\end{equation}
by which, (9.9) reduces to
\begin{equation}
Y_0 = (\gamma_{I} + \gamma_{II}) \frac{[c_{1I} X_I \mp c_{2I} X_I^{-1}]}{[c_{1I} X_I
\pm c_{2I} X_I^{-1}]} \,\,.
\end{equation}
Also from (4.10) one finds that (9.11) implies
\begin{equation}
\gamma_{I} + \gamma_{II} =0 \,\,\, \rightarrow \,\,\,  Y_0 =0 \,\,
\end{equation}
provided $m$ (or $\alpha$) takes the same value at both regions of $\Sigma_2$, and (9.17) in turn implies
(by (9.8) and (9.10)) that
\begin{equation}
C^{(p+1)}_{1} = C^{(p+1)}_{2} =0 \,\,.
\end{equation}
This is interesting. If $\Lambda$ couples with the dilaton in the
same form at both regions of $\Sigma_2$ (but see also the case I
of the next section), we do not need to introduce the brane B in
order to satisfy the matching conditions at $r=r_B$. They are
automatically satisfied as long as (9.11) and (9.15) are met.

\setcounter{equation}{0}
\section{Brane world models}

In this section we will consider various types of brane world models satisfying the matching conditions of
the previous section. In the followings we will restrict our discussion only to the case where
$|\gamma_{I}|=|\gamma_{II}| $ in consideration of symmetry. Also we will assume that the value of
$\Lambda$ of the region I ($\equiv \Lambda_I$) can be different from that of the region II ($\equiv \Lambda_{II}$).

\vskip 0.5cm
\begin{center}
{\bf (a) case I ($\bf \Lambda_{I} \neq 0$, $\bf \Lambda_{II} \neq
0$)}
\end{center}

Since $\gamma$ must be positive in the case $\Lambda \neq 0$,
$\gamma_{I}$ and $\gamma_{II}$ are both positive constants, i.e.,
$\gamma_{I}=\gamma_{II}$, which agrees with (9.11) only if
$m_{I}=- m_{II}$, where $m_{I, II}$ represent the values of $m$ at
region I and region II, respectively. Having different $m$ at each
region means that $\Lambda$ couples with dilaton differently at
each region. As an example, consider a case where $\alpha =-2$
($m=2$) in the region I, while $\alpha=0$ ($m=-2$ assuming that
$p=3$) in the region II. This describes a model in which $\Lambda$
couples with dilaton with a factor $e^{-2\Phi}$ in the region I,
while it does not couple with dilaton in the region II. Since
$m_{I}=2$ and $m_{II}=-2$, $\gamma_{I, II}$ become
$\gamma_{I}=-\alpha_{\xi I}$ and  $\gamma_{II}=\alpha_{\xi II}$,
respectively, and (9.11) implies $\gamma_{I}=\gamma_{II}$.

The analysis of Sec.9 was based on the assumption that $\alpha_I =
\alpha_{II}=\alpha$ ($m_I = m_{II}=m$). If $\alpha_I \neq
\alpha_{II}$ ($m_I \neq m_{II}$), $\mathcal{B}_1$ and
$\mathcal{B}_2$ in (9.8) must be modified. Upon using (9.15), they
are modified to
 \begin{equation}
 \mathcal{B}_1 = \kappa^2 C^{(p+1)}_1 + 2 \Big(\frac{\gamma_I}{m_I} + \frac{\gamma_{II}}{m_{II}} \Big) \frac{[c_{1I} X_I \mp c_{2I} X_I^{-1}]}{[c_{1I} X_I
\pm c_{2I} X_I^{-1}]} \,\,,
 \end{equation}
 \begin{equation}
 \mathcal{B}_2 = \kappa^2 C^{(p+1)}_2 + \frac{1}{2} \Big[(\alpha_I +2)\frac{\gamma_I}{m_I} + (\alpha_{II} +2)\frac{\gamma_{II}}{m_{II}} \Big]
 \frac{[c_{1I} X_I \mp c_{2I} X_I^{-1}]}{[c_{1I} X_I
\pm c_{2I} X_I^{-1}]} \,\,.
 \end{equation}
In the case $m_I = -m_{II}$ (and $\gamma_I = \gamma_{II} $),
(10.1) reduces to $\mathcal{B}_1 = \kappa^2 C^{(p+1)}_1$, and
consequently $\mathcal{B}_1 =0$ implies
\begin{equation}
 C^{(p+1)}_1 =0 \,\,\,\, \rightarrow V_{p+1} (\Phi)=0 \,\,.
\end{equation}
So the brane B is unnecessary even in this case. Since (10.3) also
implies $C^{(p+1)}_2 =0$, the condition $\mathcal{B}_2 =0$
requires
\begin{equation}
\Big[\frac{(\alpha_I +2)}{m_I}+ \frac{(\alpha_{II} +2)}{m_{II}} \Big] (X_I \mp X^{-1}_I) =0 \,\,,
\end{equation}
where we have set $\gamma_{I}=\gamma_{II}$ and $c_{1I}=c_{2I}=1$. For $\alpha_I =-2$ ($m_I=2$) and $\alpha_{II} =0$
($m_{I}=-2$), (10.4) is satisfied only when $X_I = X^{-1}_I$ (and therefore when $\Lambda >0$), or equivalently when
\begin{equation}
r_{0I}=r_{0II}=r_B \,\,,
\end{equation}
which together with (9.14) implies (see the definition of $d_M$ in (6.4))
\begin{equation}
d_{MI}=d_{MII}=1 \,\,.
\end{equation}
Thus in this case the integration constants $\tilde{r}_0$ and
$r_0$ (and therefore $d_M$) are all fixed by the matching
conditions. Finally, the geometry of the bulk spacetime is
described by (6.10) (or (6.14) in the case $\alpha=-2$) at both
regions of $\Sigma_2$.

\vskip 0.5cm
\begin{center}
{\bf (b) case II ($\bf \Lambda_{I} \neq 0$, $\bf \Lambda_{II} = 0$)}
\end{center}

In the case II (and in the case III) $m_I$ and $m_{II}$ do not
have to be different from each other, and in the followings we
will restrict our discussion only to the case $m_I =m_{II} =2$.
Since $m_I =m_{II}$, the whole discussion of Sec.9 can be applied
to the case II (and the case III). The brane B is unnecessary, and
the matching conditions to be met are those in (9.11), (9.14) and
(9.15). Since $m_I =m_{II}$, (9.11) implies
$\gamma_{II}=-\gamma_{I}<0$, which is allowed only if
$\Lambda_{II}=0$. (9.15) therefore reduces to
\begin{equation}
b_{0I}[X_I \pm X^{-1}_I] = b_{0II} X^{-1}_{II} \,\,,
\end{equation}
where $b_{0I}$ is given by $b_{0I} = \pm (4\gamma^2_I /
\Lambda_I)$, but $b_{0II}$ is arbitrary. So (10.7) can be
satisfied for any $r_{0I, II}$ if $b_{0II}$ is chosen properly.
Since $r_{0I, II}$ are arbitrary, $d_{MI, II}$ are not fixed by
the definition of $d_M$ in (6.4). The geometry of the bulk
spacetime is given by (6.14) in the region I, while it is given by
(6.15) in the region II.

\vskip 0.5cm
\begin{center}
{\bf (c) case III ($\bf \Lambda_{I} = \Lambda_{II} = 0$)}
\end{center}

In the case III, we only need to consider the case $m_I = m_{II} =
2$ (recall that we have decided to take $\alpha=-2$ ($m=2$) in the
case $\Lambda =0$, see Sec.6). So the whole discussion of Sec.9
can be applied to the case III either; the brane B is unnecessary,
and the matching conditions to be met are just (9.11), (9.14) and
(9.15) as before. But in the case III, (9.15) reduces to
\begin{equation}
\frac{b_{0I}}{X_I} = \frac{b_{0II}}{X_{II}} \,\,,
\end{equation}
where $b_{0I}$ and $b_{0II}$ are both arbitrary. Since $b_{0I,
II}$ are arbitrary, (10.8) can be satisfied for any arbitrary
$r_{0I, II}$, and therefore $d_{MI,II}$ are not fixed by the
matching conditions. (9.11), on the other hand, implies that
$\gamma_{I}=-\gamma_{II}>0$, so $\Sigma_2$ is closed at $\eta =0$
($r=0$) (note that the bulk geometry of the case III is given by
(6.15)), while it diverges at $\eta=\eta_m$ ($r=r_m$). Thus the
brane at $\eta =0$ is a codimension-2 brane, while the one at
$\eta=\eta_m$ is necessarily a codimension-1 brane. If we identify
these two branes as the $\eta_c \rightarrow 0$ limits of the type
II codimension-1 brane, they become T-duals\footnote{To be
precise, they are "anti T-duals" (rather than T-duals) of each
other in the sense that their masses are equal in magnitude, but
opposite in sign (see (9.11)).}  of each other because they have
the same mass (note that $|\gamma_{I}|=|\gamma_{II}|$ in the
above).

So far we have considered three types of brane world models
satisfying the matching conditions, and in all these three cases
the brane B at $r=r_B$ is unnecessary. Among these, the case I and
case II are somewhat special in the sense that $\Lambda$ or its
dilaton coupling is not uniform in the whole regions of
$\Sigma_2$. Besides this, the self-tuning of $V_p (\Phi)$ is not
obvious in the case I due to lack of extra degrees of freedom
which is needed to avoid the fine-tuning of $V_p (\Phi)$ (see
Sec.9). ( Also in the case I, (10.6) is not consistent with
$d_{\Phi} \rightarrow 0$.) Of the most interest is the case III,
which is also natural in the context of string theory where
$\Lambda$ is absent in ordinary circumstances. In the next we will
be back to the case III to go into more details about the brane
world model with $\Lambda=0$.

\vskip 0.5cm
\begin{center}
{\bf (d) case III again}
\end{center}

The self-tuning brane world model with $\Lambda=0$ includes two
coincident branes, one of which is a codimension-2 brane placed at
$\vec{\eta}=0$ (the origin of $\Sigma_2$), another a codimension-1
brane placed at $\eta =\eta_m$ (the edge of $\Sigma_2$). These two
branes are (anti) T-duals of each other, and related by the
duality relation (8.3). So the codimension-2 brane in the
$\eta-$coordinates becomes a codimension-1 brane in the
$\tilde{\eta}-$coordinates, and {\it{vice versa}}. Namely, these
two branes interchange their shapes under (8.3), and one of them
is identified with our SM-brane (plus the background NS-brane).

Let us turn to the geometry of $\Sigma_2$ especially in the
vicinity of $\vec{\eta}=0$. Since the bulk geometry is practically
unaffected by the SM-brane (see Sec.7), we simply consider the
case where $V_D (\Phi)$ is "turned off" and there is only a
background NS-brane at $\vec{\eta}=0$. With $V_D (\Phi)$ turned
off ($l=0$), (8.1) can be written as
\begin{equation}
ds^2_2 = d\eta^2 + \beta^2 \eta^2 d\theta^2 + \big(-dt^2 +
d\vec{x}^2_p \big) \,\,,~~~~~\Big(\beta \equiv
\frac{1}{d_{\xi}}\frac{r_0}{\eta_0}\Big)\,\,,
\end{equation}
where $\beta$ is a dimensionless constant associated with a
deficit angle $\delta$ defined by $\delta = 2\pi (1-\beta)$.
According to the string theoretical description (6.19), $\beta$
becomes $\beta =g^2_s \alpha_\xi$ and further, since
$\alpha_{\xi}=\kappa^2 V_p (\Phi)/2 \pi$ (note that $V_p (\Phi) =
(-1/2) \partial V_p (\Phi) /\partial \Phi$ for $l=0$), and $V_p
(\Phi) \,(= V_{NS} (\Phi))\, \sim 1/{\alpha^{\prime}}^{(p+1)/2}
g_s^2$ in the string theory, $\beta$ finally becomes
\begin{equation}
\beta \sim \Bigg(\frac{M_s}{M_{3+p}} \Bigg)^{1+p} \,\,,
\end{equation}
where $M_s$ is the string scale : $M_s =1/l_s
=1/\sqrt{\alpha^{\prime}}$. If $\beta=1$, $\Sigma_2$ is simply
$R_2$. But if $\beta=1/n$, $\Sigma_2$ becomes an orbifold $R_2
/Z_n$ with an orbifold singularity at $\vec{\eta}=0$. But in both
cases $\beta$ should be of order one; $\beta \sim \Large{O}(1)$,
which is naturally connected with the hierarchy problem. Namely in
(10.10), $M_s$ and $M_{3+p}$ should be of the same order in order
that $\beta \sim \Large{O}(1)$, which accords with an assumption
\cite{1,2,3} that there exists only one fundamental short distance
scale (i.e., the electroweak scale $m_{EW}$) in nature.

\setcounter{equation}{0}
\section{4d planck scale}

In this section we will restrict our discussion to the case $p=3$
to evaluate the 4d Planck scale $M_{pl}$. The finiteness of
$M_{pl}$ is closely related with the localization of the zero mode
of the 4d graviton. We will also concentrate our attention mostly
on the case III among the three cases of Sec.10. Finally we will
use $l \cong 0$ by the same reason that was used to obtain
(10.10).

In the case III, the 6d metric is given by (8.1) in the region I
($0 \leq \eta \leq \eta_B$), while it is given by (8.2) with
$\eta$ replaced by $\eta_m - \eta$ in the region II ($\eta_B <
\eta \leq \eta_m$) (also $\eta_0$ must be identified with $\eta_B$
in order that two metrics match each other at $\eta =\eta_B$). For
this metric, $M_{pl}$ is given by
\begin{equation}
M_{pl}^2 = 2 \pi \, M{_6^4} \, \frac{|\alpha_{\xi}|}{d_B}\, \Bigg( \int_{\tau}^{\eta_B} \eta d\eta + \eta_B^4
\int^{\eta_m -\tau}_{\eta_B} \frac{d\eta}{(\eta_m - \eta)^3} \Bigg) \,\,,
\end{equation}
where we have used (6.17) (together with (6.13)) and the relation
$d_\xi =d^{-2}_{\Phi} d^2_B$. In (11.1), $\tau$ represents the
thickness (we assume that every brane has the same thickness) of
the branes in the $\eta-$coordinates. In the thin brane limit,
$\tau$ vanishes. But in reality, branes have nonzero thickness and
$\tau$ takes some nonzero value. Neglecting $\tau^2$ term, one
finds
\begin{equation}
M_{pl}^2 = 2 \pi \, M{_6^4} \, \frac{|\alpha_{\xi}|}{d_B}\,
\rho^2_{max} \,\,, ~~~~~\big(\rho_{max} \equiv \frac{\eta^2_B}{\tau}\big) \,\,,
\end{equation}
where $\rho_{max}$ may be identified with the size of the
codimension-1 brane, or equivalently the size of $\Sigma_2$;
indeed, if the codimension-2 brane has a size (thickness) $\tau$,
then the codimension-1 brane must have a size $\eta^2_B /\tau$($=
\rho_{max}$) by (8.3) because they are (anti) T-duals of each
other. Since $\alpha_{\xi} = \kappa^2 V_p (\Phi)/ 2\pi$ for $l =
0$, (11.2) finally becomes
\begin{equation}
M{_{pl}^2} \sim \big|V_p (\Phi) \big| \rho^2_{max} \,\,,
\end{equation}
where we have set $d_B =1$ (note that $a_B=0$ for $l=0$ and
$\alpha=-2$). (11.3) contrasts with the conventional equation
\cite{2}
\begin{equation}
M_{pl}^2 \sim M_6^4 \, \rho_{max}^2 \,\,.
\end{equation}
In (11.4), the 4d Planck scale $M_{pl}$ is given in terms of the
6d Planck scale $M_6$. But in (11.3), $M_{pl}$ is not directly
given by $M_6$ ($M_6^4$ was cancelled out in (11.3)); it is
determined by the brane tensions and the size of the extra
dimensions. But if we use $V_p (\Phi) \sim V_{NS}(\Phi) \sim 1/
{\alpha^{\prime}}^2 g_s^2$, (11.3) can be written in terms of
$M_s$ :
\begin{equation}
M_{pl}^2 \,\, \sim \,\,  \frac{M_s^4}{g_s^2} \, \rho_{max}^2 \,\,,
\end{equation}
where $M_s$ is the string scale: $M_s = 1/l_s = 1/ \sqrt{\alpha^{\prime}}$ as mentioned before.
In some sense, (11.5) may be considered as a string theoretic generalization of (11.4);
(11.5) reduces to (11.4) if we identify
\begin{equation}
M_6^4 =  \frac{M_s^4}{g_s^2} \,\,,
\end{equation}
which is just an analogue of the string theoretical definition of
the 10d Planck scale: $M_{10}^8 = M_s^8 / g_s^2$. In the limit
$g_s \rightarrow 0$, however, (11.6) is unnatural because it
implies $M_{6}/M_{s} \rightarrow \infty$ as $g_s \rightarrow 0$.
(11.6) may be senseful perhaps when $g_s \sim \large{O}(1)$. That
is, the identification of (11.5) with (11.4) may not be valid in
the weakly coupled case $g_s \rightarrow 0$.

There is a different way of viewing (11.5). If we take an
assumption that $m_{EW}$ is the only fundamental short distance
scale in nature (i.e., if we assume that $M_s \sim TeV$ and
$\rho_{max} \sim TeV^{-1}$), then $g_s$ expected from (11.5) would
be $\sim 10^{-16}$, which is just the realistic decoupling limit
considered in "Little String Theories at a $TeV$"\cite{20}.

Finally, $M_{pl}$ of the case II (of Sec.10) is similar to that of
the case III. For $m_I =m_{II}=2$ ($\alpha_I =\alpha_{II}=-2$), it
is given by
\begin{equation}
M_{pl}^2 = 2 \pi \, M{_6^4} \, \frac{|\alpha_{\xi}|}{d_B}\, \Bigg( \frac{1}{\sqrt{|\Lambda|}}\int_{\tau}^{\eta_B}
\sinh{\sqrt{|\Lambda|}}\eta d\eta + \eta_B^4 \int^{\eta_m -\tau}_{\eta_B}
\frac{d\eta}{(\eta_m - \eta)^3} \Bigg) \,\,.
\end{equation}
Note that (11.7) reduces to (11.1) in the limit $\Lambda
\rightarrow 0$.  For the case I, we will not give a precise value
of $M_{pl}$, but it is obviously finite even in the thin brane
limit $\tau \rightarrow 0$. Omitting $\eta_B -$dependent
hyperbolic functions, it is of order $M_{pl}^2 \sim |V_p
(\Phi)|/|\Lambda|$, or in terms $M_s$ it is given by $M_{pl}^2
\sim M_s^4 / |\Lambda| g_s^2$. But as mentioned before, the
self-tuning is not guaranteed in this case.

\setcounter{equation}{0}
\section{Summary}
In this paper we have presented a new type of self-tuning
mechanism for $(3+p)$d brane world models in the framework of
gravity-scalar theory. Each model contains two coincident branes
each of which is a stack of a D-brane (SM-brane) and a background
NS-brane. Among these models, of the most interest is the case
with $\Lambda=0$, which is not only interesting but also natural
in the context of the string theory. In this model, one of the
coincident branes is a codimension-2 brane placed at the origin
$\vec{\eta}=0$ of the 2d transverse space $\Sigma_2$, while the
other is a codimension-1 brane placed at the edge of $\Sigma_2$.
These two branes are (anti) T-duals of each other, and interchange
their shapes under duality transformation, and one of them is
identified as our SM-brane (plus the background NS-brane).

The given models exhibit a remarkable feature. In the limit $g_s
\rightarrow 0$, the bulk geometry (as well as the flat intrinsic
geometry of the branes) is practically insensitive to the quantum
fluctuations of SM-fields with support on the SM-brane. This can
be achieved by introducing NS-brane which serves as a background
brane on which our SM-brane is to be set. Indeed, the existence of
the background NS-brane is naturally suggested by field equations,
which impose a certain restriction on (the dilaton coupling of)
the brane tension so that the background brane must be of the
NS-NS type. In the presence of this NS-brane the 2d transverse
space $\Sigma_2$ becomes an orbifold $R_2 /Z_n$ with a deficit
angle $\delta = 2\pi (1-\beta)$ where $\beta \sim (M_s
/M_{3+p})^{1+p}$. So in order that $\beta$ becomes of order one,
the $(3+p)$d Planck scale $M_{3+p}$ should be of the same order as
the string scale $M_s$, which accords with the hierarchy
conjecture that there may exist only one fundamental short
distance scale in nature.

Now introduce the SM-brane on top of the background NS-brane
placed at the orbifold singularity. Such an introduction of an
SM-brane usually affects the geometry of bulk spacetime due to the
tension of the SM-brane. In our case the effect of the (brane
tension of the) SM-brane on the bulk geometry is essentially
expressed in terms of the parameters $l$ and $k$. So they take
nonzero values in the presence of the SM-brane, while they vanish
in the absence of the SM-brane. In the given models, however, the
increments in $l$ and $k$ due to an introduction of the SM-brane
is negligibly small; they are all of the order $\sim g_s$, so they
go to zero in the limit $g_s \rightarrow 0$. The shifts in $l$ and
$k$ due to quantum corrections to the brane tension (due to
dynamics of SM fields on the SM-brane) are consequently even
smaller than this. They are just $\delta l, \delta k$ $ \sim
 g^2_{\mathrm YM} g_s \, {\alpha^{\prime}}^{-(p-3)/2}$ (or $\delta l,
\delta k$ $ \sim g^4_{\mathrm YM} \, {\alpha^{\prime}}^{-(p-3)}
\sim g_s^2$ in the case $g^2_{\mathrm YM} \sim g_s
{\alpha^{\prime}}^{-(p-3)/2}$), which again vanish in the limit
$g_s \rightarrow 0$. Thus the geometry of the bulk spacetime is
practically insensitive to the quantum fluctuations of SM fields
with support on the SM-brane.

Finally for $p=3$, the 4d Planck scale $M_{pl}$ is given (in the
case $\Lambda =0$) by $M_{pl}^2 \sim M_s^4 \rho_{max}^2 /g_s^2$.
So if we apply to this equation the hierarchy assumption that
$M_s$ and $\rho_{max}$ are both of the order the electroweak scale
$\sim TeV$, the estimated value for $g_s$ will be about $\sim
10^{-16}$, which, however, is just the realistic value of the
decoupling limit $g_s \rightarrow 0$. This again shows that taking
$g_s \rightarrow 0$ is consistent with the hierarchy conjecture in
the present paper.

\vskip 1cm
\begin{center}
{\large \bf Acknowledgement}
\end{center}

This work was supported by Kyungsung University in 2005.

\vskip 1cm
\renewcommand{\theequation}{A.\arabic{equation}}
\begin{center}
{\large \bf Appendix : Codimension-1 branes}
\end{center}
\setcounter{equation}{0}

In this section we will consider two different types of
codimension-1 branes, which we will call type I or type II
codimension-1 brane, respectively.

\vskip 0.5cm
\begin{center}
{\bf A.  Type I}
\end{center}

Type I codimension-1 brane (this is an ordinary codimension-1
brane) is described by
\begin{equation}
I_{cod-1}^{(I)} = - \int d^{p+1}x d\theta
\sqrt{-\det{|g_{\mu\nu}|}} \,\sqrt{g_{\theta\theta}}\,
V_{p+1}({\Phi}) \,\,,
\end{equation}
where $g_{\theta\theta}$ (as well as $g_{\mu\nu}$) is a pullback
of $G_{\theta\theta}$ to the codimension-1 brane. Upon using (2.2)
and (2.6), (A.1) reduces to
\begin{equation}
I_{cod-1}^{(I)} = - \int_{\Sigma_3} d^3 y \sqrt{\hat{g}_{tt}}
\,\sqrt{\hat{g}_{\theta\theta}}\,\,e^{4\Phi -\frac{p}{2}B} V_{p+1}
({\Phi})\, \delta (r-r_B ) \,\,,
\end{equation}
where $r_B$ represents the position of the codimension-1 brane.
With this action, the field equations become
\begin{equation}
\frac{d}{dr}\Big(r \frac{d\ln R}{dr} \Big) + \Lambda \psi r = -
\kappa^2 C_1^{(p+1)} \,\delta (r-r_B) \,\,,
\end{equation}
\begin{eqnarray}
\frac{d}{dr}\Big(r \frac{d\Phi}{dr} \Big) - \frac{[(\alpha
+2)(p+1)+4]}{8}\Lambda \psi r = \frac{(p+1)}{2}\kappa^2
C_2^{(p+1)}\, \delta (r-r_B) \nonumber \\+ \frac{1}{2} \kappa^2
C_1^{(p+1)} \, \delta (r-r_B) \,\,,~~~~~~~~~~
\end{eqnarray}
\begin{equation}
\frac{d}{dr}\Big(r \frac{dB}{dr} \Big) - \frac{(\alpha
+2)}{2}\Lambda \psi r = 2 \kappa^2 C_2^{(p+1)} \delta (r-r_B) \,\,,
\end{equation}
where the constants $C_{i}^{(p+1)}$ are defined by
\begin{eqnarray}
C_{1}^{(p+1)} = e^{(p+1)B/2} \sqrt{g_{\theta\theta}}\, V_{p+1}
(\Phi) \Big|_{r=r_B} \,\,, \nonumber \\ C_{2}^{(p+1)} =
e^{(p+1)B/2} \sqrt{g_{\theta\theta}} \Big( V_{p+1} (\Phi) +
\frac{1}{2} \frac{\partial V_{p+1} (\Phi)}{\partial \Phi}\Big)
\Big|_{r=r_B} \,\,.
\end{eqnarray}

\vskip 0.5cm
\begin{center}
{\bf B.  Type II}
\end{center}

Type II codimension-1 brane can be obtained from (A.1) (or
(A.2)) by setting
\begin{equation}
\sqrt{g_{\theta\theta}}\,V_{p+1} (\Phi) \equiv \frac{V_p
(\Phi)}{2\pi} \,\,.
\end{equation}
So the action is given by
\begin{equation}
I^{(II)}_{cod-1} = - \int_{\Sigma_3} d^3 y \sqrt{-\hat{g}_{tt}} \,
e^{2\Phi} \, V_p (\Phi) \frac{\delta(r)}{2\pi} \,\,,
\end{equation}
where we have assumed that $\Sigma_2$ is not closed at $r=0$ and
the brane is placed there. (A.7) indicates that $V_{p+1} (\Phi)$
of the type II brane is inversely proportional to the size (the
radius) of the brane; i.e., $V_{p+1} (\Phi) \propto
1/\sqrt{g_{\theta\theta}}$ so that $V_p (\Phi)$ becomes
independent of $\sqrt{g_{\theta\theta}}$. This is the crucial
difference between type I and Type II codimension-1 branes. In the
case of type I brane, $V_{p+1} (\Phi)$ is itself independent of
$\sqrt{g_{\theta\theta}}$. (A.8) has the same form as (3.3) except
that the 2d delta-function $\delta^2 (\vec{r})$ is replaced by 1d
delta-function $\delta(r)/2\pi \, \sqrt{\hat{g}_2}$. The type II
codimension-1 brane is a codimension-1 brane which can be obtained
from a codimension-2 brane (a point) by expanding it to a circle
with the total mass of the brane kept constant. Conversely, a
codimension-2 brane can be obtained from the type II codimension-1
brane by shrinking it to a point with the total mass of the brane
kept constant. (A.7) indicates the fact that the total mass of the
brane is unchanged under this expansion (or contraction). The
field equations for this codimension-1 brane are given by
\begin{equation}
\frac{d}{dr}\Big(|r| \frac{d\ln R}{dr} \Big) + \Lambda \psi |r|= -
2 \kappa^2 C_1 \,\frac{\delta(r)}{2\pi} \,\,,
\end{equation}
\begin{equation}
\frac{d}{dr}\Big(|r| \frac{d\Phi}{dr} \Big) - \frac{[(\alpha
+2)(p+1)+4]}{8}\Lambda \psi |r| = (p+1)\kappa^2 C_2
\frac{\delta(r)}{2\pi} \,\,,
\end{equation}
\begin{equation}
\frac{d}{dr}\Big(|r| \frac{dB}{dr} \Big) - \frac{(\alpha
+2)}{2}\Lambda \psi |r| = 4 \kappa^2 C_2 \frac{\delta(r)}{2\pi}
\,\,,
\end{equation}
where the constants $C_{i}$ are defined by
\begin{equation}
C_{1} = e^{(p+1)B/2} V_{p} (\Phi) \Big|_{r=0} \,\,, ~~~~~ C_{2} =
e^{(p+1)B/2} \Big( V_{p} (\Phi) + \frac{1}{2} \frac{\partial V_{p}
(\Phi)}{\partial \Phi}\Big) \Big|_{r=0} \,\,,
\end{equation}
similarly to (3.26). Note that the right hand sides of
(A.9)-(A.11) have been doubled as compared with (3.21)-(3.23). The
reason is as follows. The left hand sides of (A.9)-(A.11) show
that there is a reflection symmetry about the codimension-1 brane
at $r=0$ (we have $|r|$ in the equations). So we naturally take
the codimension-1 brane as an orbifold fixed line: i.e., we
identify every point of the region $r<0$ with the corresponding
point of the region $r>0$, and then take the region $r>0$ as a
fundamental domain. Then the orbifold fixed line at $r=0$ becomes
a stack of two codimension-1 branes each of which belongs to the
corresponding regions. So we have to double the right hand sides
of (A.9)-(A.11). The equations (A.9)-(A.11) can be solved by (4.1)
with $\psi (r)$ and $i_M (r)$ given by
\begin{equation}
\frac{d}{dr} \Big(|r| \frac{d\ln \psi r^2}{dr} \Big) + \frac{m
\Lambda}{|r|} \psi r^2 = - 2 m \alpha_{\xi} \delta (r)
\end{equation}
and
\begin{equation}
\frac{d}{dr} \Big( |r| \frac{d\ln {\hat{i}}_{M}}{dr}\Big) = 2 \alpha_{M} \delta
(r)\,\,, ~~~( \hat{i}_{M} \equiv i_M r^{-2k_M} )
\end{equation}
where $\alpha_M$ are the same ones as those in (4.4)-(4.7).

\end{document}